\documentclass[prd,onecolumn,showpacs,preprintnumbers,amsmath,amssymb,color]{revtex4}
\usepackage{graphicx}
\usepackage{dcolumn}
\usepackage{bm}
\usepackage{color}
\definecolor{cyan}{cmyk}{1.,0.,0.,0.5}
\definecolor{vert}{cmyk}{0.5,0.,0.5,0.5}
\definecolor{magenta}{cmyk}{0.,1.,0.,0.5}
\definecolor{verdatre}{cmyk}{0.5,0.,0.5,0.5}
\definecolor{yellow}{cmyk}{0.,0.,1.,0.0}
\definecolor{rouge}{cmyk}{0.,0.4,0.6,0.0}
\definecolor{orange}{cmyk}{0.,0.5,0.5,0.}
\definecolor{violet}{rgb}{0.5,0.,0.5}
\newcommand{\aaa}{\hspace{0.3cm}}
\def\GE73{{}^{73}{\rm Ge}}
\def\gE76{{}^{76}{\rm Ge}}
\def\gEe74{{}^{74}{\rm Ge}}
\def\xe131{{}^{131}{\rm Xe}}
\def\i127{{}^{127}{\rm I}}

\def\dj{\hbox{d\kern-0.347em \vrule width 0.3em height 1.252ex depth
-1.21ex \kern 0.051em}}

\newcommand{\be}{\begin{equation}}
\newcommand{\ee}{\end{equation}}
\newcommand{\ben}{\begin{equation*}}
\newcommand{\een}{\end{equation*}}
\newcommand{\ba}{\begin{eqnarray}}
\newcommand{\ea}{\end{eqnarray}}
\newcommand{\ban}{\begin{eqnarray*}}
\newcommand{\ean}{\end{eqnarray*}}
\newcommand{\brr}{\begin{array}}
\newcommand{\err}{\end{array}}
\newcommand{\bc}{\begin{center}}
\newcommand{\ec}{\end{center}}

\newcommand{\bea}{\begin{eqnarray}}
\newcommand{\eea}{\end{eqnarray}}
\newcommand{\bean}{\begin{eqnarray*}}
\newcommand{\eean}{\end{eqnarray*}}

\newcommand\lsim{\mathrel{\rlap{\lower4pt\hbox{\hskip1pt$\sim$}}
    \raise1pt\hbox{$<$}}}
\newcommand\gsim{\mathrel{\rlap{\lower4pt\hbox{\hskip1pt$\sim$}}
    \raise1pt\hbox{$>$}}}

\newcommand{\centeron}[2]{{\setbox0=\hbox{#1}\setbox1=\hbox{#2}\ifdim
                             \wd1>\wd0\kern.5\wd1\kern-.5\wd0\fi \copy0
                             \kern-.5\wd0\kern-.5\wd1\copy1\ifdim\wd0>\wd1
                             \kern.5\wd0\kern-.5\wd1\fi}}
\newcommand{\ltap}{\>\centeron{\raise.35ex\hbox{$<$}}
                     {\lower.65ex\hbox{$\sim$}}\>}
\newcommand{\gtap}{\>\centeron{\raise.35ex\hbox{$>$}}
                     {\lower.65ex\hbox{$\sim$}}\>}

\begin{document}
\preprint{LAPTH--1102/05}
\preprint{SPhT--T05/102}
\preprint{DFTT--16/2005}
\title{Kaluza-Klein Dark Matter and Galactic Antiprotons}
\author{Aurelien Barrau}
\email{Aurelien.Barrau@cern.ch}
\affiliation{
Laboratory for Subatomic Physics and Cosmology\\
CNRS-IN2P3 and Universit\'e Joseph Fourier\\
53, avenue des Martyrs, 38026 Grenoble cedex, France}

\author{Pierre Salati}
\email{Salati@lapp.in2p3.fr}
\affiliation{
Laboratoire d'Annecy-le-Vieux de Physique Th\'eorique\\
CNRS-SPM and Universit\'e de Savoie\\
9, Chemin de Bellevue, B.P.110 74941 Annecy-le-Vieux, France}

\author{G\'eraldine Servant}
\email{servant@spht.saclay.cea.fr}
\affiliation{
Service de Physique Th\'eorique\\
CEA Saclay\\
91191 Gif-sur-Yvette C\'edex, France}

\author{Fiorenza Donato }
\affiliation{
Dipartimento di Fisica Teorica and INFN, Torino\\
via Giuria 1, 10125 Torino, Italy}

\author{Julien Grain}
\affiliation{
Laboratory for Subatomic Physics and Cosmology \\
CNRS-IN2P3 and Universit\'e Joseph Fourier\\
53, avenue des Martyrs, 38026 Grenoble cedex, France}

\author{David Maurin}
\affiliation{
Service d'Astrophysique DSM/SAp\\
Orme des Merisiers, CEA/Saclay\\
91191 Gif-sur-Yvette Cedex, France}

\author{Richard Taillet}
\affiliation{
Universit\'e de Savoie, 73000 Chamb\'ery, France\\
LPNHE, Jussieu, CNRS-IN2P3, Paris, France}
\date{June 15, 2005}
%
\begin{abstract}
\vskip 0.2cm
Extra dimensions offer new ways to address long-standing problems in beyond
the standard model particle physics. In some classes of extra-dimensional models,
the lightest Kaluza-Klein particle is a viable dark matter candidate. In this work,
we study indirect detection of Kaluza-Klein dark matter via its annihilation into
antiprotons. We use a sophisticated galactic cosmic ray diffusion model whose
parameters are fully constrained by an extensive set of experimental data. We
discuss how fluxes of cosmic antiprotons can be used to exclude low Kaluza-Klein
masses.
\end{abstract}
\maketitle
\newpage
\section{Introduction}

Non-baryonic dark matter has been shown to be the dominant matter component
of our Universe by several independent measurements -- see \cite{turner} for a
review. The recently published WMAP results \cite{wmap}, combined with 
ACBAR, CBI and 2dFGRS, lead to precise estimates of the baryonic, matter and total
densities~:
$\Omega_b h^2 = 0.0224 \pm 0.0009$,
$\Omega_m h^2 = 0.135  \pm 0.009$ and 
$\Omega_{tot} = 1.02   \pm 0.02$.
Weakly Interacting Massive Particles (WIMPs) are the favourite candidates to account
for the Cold (non-baryonic) Dark Matter (CDM) as the required relic density can be
naturally generated. Theoretically well-founded, neutralinos are certainly the most
extensively studied example. On the other hand, in spite of the very important efforts
devoted to direct and indirect searches in this direction, supersymmetric particles
have not yet been discovered and alternative candidates should be considered.
Among them, Kaluza-Klein (KK) particles are promising. So far, they arise as stable
viable WIMPs in two frameworks~: In Universal Extra Dimensions (UED) \cite{appel}
and in some warped geometries \`a la Randall-Sundrum \cite{rs}.

\vskip 0.1cm
In the case of UED, all standard model fields propagate in one or more flat compact
extra dimensions -- unlike models with Large Extra Dimensions \`a la ADD \cite{led}.
As a result, the combination of a translation by $\pi R$ with a flip of sign of all
odd states in the KK Fourier decomposition of the bulk fields -- named KK-parity --
is conserved. This implies that the lightest first level KK particle (LKP) cannot decay
into standard model modes and is stable. Such a Kaluza-Klein particle is likely to
be associated with the first KK excitation of the photon, more precisely the first
excitation of the hypercharge gauge boson \cite{cheng2}, and is refered to as
$B^{(1)}$. Depending on the number of dimensions and on the mass difference between
the LKP and the NLKP -- next to LKP -- the $B^{(1)}$ mass is expected to lie in the
range 300 -- 1000 GeV if it is to account for dark matter \cite{servant1} -- for a recent
analysis see \cite{relic2}. Although not very narrow, this range is much smaller than
in the neutralino case and this approach has much less free parameters. Furthermore,
this range is fully compatible with experimental constraints which lead -- in the
$D=5$ case -- from precision electroweak measurements to compactification radii
satisfying $R^{-1} \gsim 300$~GeV.
Direct detection of the $B^{(1)}$ LKP has been studied in germanium, sodium iodine
and xenon detectors \cite{directLKP,cheng}. Indirect detection through gamma-rays
\cite{cheng,servant2,gammas1,gammas2,gammas3}, neutrinos and synchrotron flux
\cite{servant2}, or through positrons \cite{cheng,hooper2} has also been considered.
The neutrino spectrum from LKP annihilation in the Sun was investigated in
\cite{hooper1}. Constraints on UED models from radion cosmology have also been
studied \cite{KST}.

\vskip 0.1cm
The second class of Kaluza-Klein WIMPs arises in higher dimensional warped Grand
Unified Theories \cite{Agashe1,Agashe2}. In these models, a stable KK fermion can
arise as a consequence of imposing proton stability in a way very reminiscent to
R-parity stabilizing the lightest supersymmetric particle in supersymmetric models.
The symmetry is called $Z_3$ and the Lightest $Z_3$ Particle (LZP) is stable since
it cannot decay into standard model particles. It is actually associated with a KK
Dirac right-handed neutrino with a mass in the 1 GeV to 1 TeV range. This RH neutrino
has gauge interactions in particular with additional KK $Z^{\prime}$ gauge bosons.
Nevertheless, its interactions with ordinary matter are feeble because they involve
heavy gauge bosons with a mass $\gsim 3$ TeV. This opens the possibility of a weakly
heavy gauge bosons with a mass $M_{\rm KK} \gsim 3$ TeV. This opens the possibility
of a weakly interacting Dirac RH neutrino. Indirect detection of ``warped dark matter"
in neutrino telescopes, gamma ray telescopes and cosmic positron experiments was
investigated in \cite{LZPindirect}.
In principle, the LZP is not necessarily the lightest KK particle. There might be
lighter KK modes but which are unstable because they are not charged under $Z_3$.
In practise though, and in the models of \cite{Agashe1,Agashe2}, the RH neutrino LZP
turns out to be the lightest KK particle due to various phenomenological constraints.
Thus, in the following, we will use the generic appellation ``LKP" for both UED and
warped types of KK dark matter.

\vskip 0.1cm
In the present paper, we study the cosmic antiprotons that should be emitted
as a result of LKP annihilations in the halo of the Milky Way. Those cosmic rays
are of particular interest as the $\bar{p}/p$ ratio is both small -- smaller than
$10^{-4}$ whatever the energy -- and well-known \cite{Usine2}. The antiproton flux
has been mostly measured by stratospheric balloon borne detectors --
IMAX \cite{imax},
MASS \cite{mass_91},
CAPRICE \cite{caprice_94,caprice_98} and
BESS \cite{bess95_97,bess_98,bess_00} -- flying at the top of the atmosphere.
The interactions of high energy particles impinging on the latter generate a background
to be removed in order to measure a signal that is compatible -- given the uncertainties
-- with a pure secondary production arising from the spallations of cosmic ray nuclei
on the interstellar gas of the Milky Way disk. The antiproton flux will be measured
with unprecedented accuracy by the forthcoming space experiment AMS \cite{ams} that
has already flown on the space shuttle \cite{ams_98}. Small deviations from a pure
secondary energy spectrum -- expected in our case if LKP particles annihilate in the
galactic halo -- are potentially detectable by AMS.
In section~II, the source term is computed by convolving the LKP number density and
cross sections with the relevant fragmentation functions.
Section~III is devoted to the propagation scheme and the astrophysical parameters.
Finally, the primary antiproton flux resulting from LKP annihilations is compared
in section~IV with the secondary background and some perspectives are drawn. We will
show that antiprotons can at least constrain the lowest values of LKP masses.

\newpage
\section{Source term}

The production rate $q_{\bar{p}}^{\rm LKP}$ of antiprotons is obtained from
the convolution over the various annihilation channels $f$ of the appropriate
annihilation cross section $<\sigma v>_{f}$ with the fragmentation function
$\left( {dN_{\bar{p}}}/{dT_{\bar{p}}} \right)_{f}$. It can be written as~:
\be
q_{\bar{p}}^{\rm LKP} \left( r , T_{\bar{p}} \right) \; = \;
{\displaystyle \frac{1}{2}} \,
{\displaystyle \sum_{{\rm channel} \, f}} \, <\sigma v>_{f} \,
\left( {\displaystyle \frac{dN_{\bar{p}}}{dT_{\bar{p}}}} \right)_{f} \;
\left\{
n_{\rm LKP}(r) \equiv {\displaystyle \frac{\rho_{\rm LKP}(r)}{M_{\rm LKP}}}
\right\}^2 \,\, .
\label{source_term}
\ee
The LKP particles in the initial state are identical hence the
overall factor of $1/2$. The distance between the production point and
the galactic center is denoted by $r$ while $n_{\rm LKP}(r)$ and $\rho_{\rm LKP}(r)$
respectively stand for the LKP number and mass densities. Antiprotons are
produced with a kinetic energy $T_{\bar{p}}$ that ranges from 0 up to the LKP
mass $M_{\rm LKP}$. The previous relation features the four key ingredients that
participate into the Kaluza-Klein antiproton souce term~$q_{\bar{p}}^{\rm LKP}$.

\vskip 0.1cm
To commence, for UED dark matter, the LKP annihilation cross section into
fermions is given in the non-relativistic expansion limit by \cite{servant1}~:
\be
\sigma \left| \vec{v_{1}} - \vec{v_{2}} \right|
\left\{ B^{(1)} B^{(1)} \rightarrow f \bar{f} \right\} \; = \;
{\displaystyle \frac{8 \pi}{9}} \, N_{c} \,
\left( Y_{L}^{4} \, + \, Y_{R}{^4} \right) \,
{\displaystyle \frac{\alpha_{1}^{2}}{M_{\rm LKP}^{2}}} \,
\left( 1 - v^{2} \right) \,\, ,
\label{cs_UED}
\ee
where $N_{c}$, $Y_{L}$ and $Y_{R}$ are respectively the number of colors and the
left and right hypercharges of the resulting fermion $f$ whereas
$2 \vec{v} \, = \, \vec{v_{1}} -  \vec{v_{2}}$. Velocities within the Milky Way are
typically non-relativistic so that the factor $1 - v^{2}$ can safely be approximated
by 1. Notice that in contrast with neutralino dark matter, the annihilation into
fermions is not helicity-suppressed.
For warped dark matter, there is no simple analytical formula -- in particular,
couplings depend in a non-trivial way on the LKP mass -- but we can summarize the
situation as follows. For LKPs lighter than approximately 100 GeV, LKP annihilations
proceed dominantly via s-channel $Z$-exchange. For larger masses, annihilation into
top quarks via the t-channel exchange of the GUT KK gauge boson $X_{s}$ or into
$t \overline{t}$, $b \overline{b}$, $W^{+} W^{-}$ and $Z h$  via the s-channel KK
$Z^{\prime}$ exchange dominates. We refer the reader to the appendices of \cite{Agashe2}
for details. In the rest of the paper, we have taken a Kaluza-Klein gauge boson mass
of $M_{\rm KK} = 3$ TeV and varied the LZP mass from 30 to 70 GeV. Median values for
the annihilation cross section have been assumed here. They correspond to the couplings
$g_{10} = (g' + g_{s}) / 2 = 0.785$ and $c_{\nu'_{L}} = 0.4$.

\vskip 0.1cm
The second ingredient is the computation of the number of antiprotons with kinetic
energy between $T_{\bar{p}}$ and $T_{\bar{p}}+dT_{\bar{p}}$ formed within a jet
induced by a $q \bar{q}$ pair of energy $M_{\rm LKP}$. This was evaluated with the
high-energy physics frequently-used Monte-Carlo event generator {\sc pythia} \cite{tj},
based on the so-called string fragmentation model.

\vskip 0.1cm
The square of the LKP number density $n_{\rm LKP}$ enters into the annihilation
rate~(\ref{source_term}) and scales as $M_{\rm LKP}^{-2}$. In UED models, the LKP
requires a mass in the range between 700 and 900 GeV in order to generate the
observed thermal relic density, unless other KK modes participate in the freeze-out
process \cite{servant1}. If this is the case, somewhat smaller masses are possible.
A recent analysis, taking into account the effects of second level KK modes, indicates
that the upper edge of this mass range is favored \cite{relic2}. In any case, we should
keep in mind that the precise prediction of the LKP relic density depends on the
particular KK mass spectrum which is used and is somewhat model-dependent. We will
therefore consider masses in the lower range $\sim 300$ GeV which is the most favorable
case as far as the antiproton signal is concerned.
As for the warped LKP, it can thermally generate the observed quantity of dark matter
in two mass ranges~: near the $Z$-resonance with $M_{\rm LZP} \approx$ 20 to 80 GeV
and for considerably heavier masses -- $M_{\rm LZP} \gsim$ several hundred GeV --
\cite{Agashe1,Agashe2}. Again, we will restrict ourselves to the more easily accessible
lowest masses.

\vskip 0.1cm
The distribution of dark matter inside galaxies is still an open and very
debated issue.
From one side, results from cosmological N-body simulations in $\Lambda$-CDM models
\cite{nfw,fukushige,moore} indicate a universal and coreless dark matter density
profile. At small radii, the latter diverges with the distance $r$ from the galactic
center as $r^{- \gamma}$ with $\gamma \sim$ 1 to 1.5. This implies a strongly peaked
dark matter density at galactic centers. Very recent results obtained from simulations
of halo formation \cite{2004MNRAS.349.1039N, 2004MNRAS.353..624D} strongly disfavour
a singularity as steep as 1.5 and seem to point toward slopes logarithmically
dependent on the distance from the galactic center and no steeper than $\sim$ 1.2.
It should indeed be noticed that these cusps are predicted in regions which are usually
smaller than the typical resolution size of the simulations.
On the other side, several analysis of rotational curves observed for galaxies of
different morphological types
\cite{bs,deblok,swaters,weldrake,2004MNRAS.351..903G,2004MNRAS.353L..17D}
put serious doubts on the existence of dark matter cusps in the central regions
of the considered objects. Instead of a central singularity, these studies rather
suggest a cored dark matter distribution, flattened toward the central regions.
In the present analysis, we will consider the generic dark matter distribution
\be
\rho_{\rm CDM}(r) \; = \; \rho_{{\rm CDM} \; \odot} \,
\left\{
{\displaystyle \frac{r_{\odot}}{r}} \right\}^{\gamma} \,
\left\{
{\displaystyle \frac{1 \, + \, \left( r_{\odot} / a \right)^{\alpha}}
{1 \, + \, \left( r / a \right)^{\alpha}}}
\right\}^{\left( \beta - \gamma \right) / \alpha} \;\; ,
\label{eq:profile}
\ee
where $r_{\odot} = 8$ kpc is the distance of the Solar System from the galactic
center. The local -- Solar System -- CDM density has been set equal to
$\rho_{{\rm CDM} \; \odot} = 0.3$ GeV cm$^{-3}$. In the case of the
pseudo-isothermal profile, the typical length scale $a$ is the radius of
the central core. The profile indices $\alpha$, $\beta$ and $\gamma$ for
the dark matter distributions which we have considered here are indicated in
Tab.~\ref{tab:indices}.
%
\begin{table}[h!]
\begin{center}
{\begin{tabular}{@{}|l|c|c|c|c|@{}}
\hline
Halo model & $\alpha$ & $\beta$ & $\gamma$ & $a$ [kpc] \\
\hline
\hline
Cored isothermal~\cite{bahcall}
& {\aaa} 2 {\aaa} & {\aaa} 2 {\aaa} & {\aaa} 0 {\aaa} & {\aaa} 4 
{\aaa} \\
Navarro, Frenk \& White~\cite{nfw}
&        1        &        3        &        1        &        
25       \\
Moore~\cite{moore}
&        1.5      &        3        &        1.3      &        
30       \\
\hline
\end{tabular}}
\end{center}
\caption{
Dark matter distribution profiles in the Milky Way.
\label{tab:indices}}
\end{table}
%
As already underlined in \cite{MaurinTaillet,pbar_susy} -- and as it will be
clear also from the results presented in the following of the present paper --
the diffusion of primary CDM generated antiprotons is only very mildly dependent
on the chosen dark matter distribution function.

\section{Galactic propagation: control of uncertainties}

Propagation in the Galaxy, while studied for a long time is not a
simple matter. A realistic description should take into account the
coupling between gas, magnetic field and cosmic rays (CRs). This is
far from being reached -- at least at the Galactic scale. Our lack of
knowledge about the structure of magnetic turbulences and their spatial
distribution -- probably related to the regions of star formation --
hampers any clear and unambiguous description of the transport of CRs.
So far, one major approximation assumed in all -- but a very few number --
of papers is that diffusion in the Galaxy does not depend on the galactic
position. Even with this simplification, transport of CRs is not straightforward.
It involves the now classical following ingredients~: diffusion -- random walk
on magnetic inhomogeneities -- and convection -- directed outward the galactic
disk -- which compete for the spatial transport, especially at low energy.
Regarding the energetic balance, energy losses -- Coulomb, ionization and
adiabatic -- replenish the low energy tail whereas momentum diffusion --
reacceleration -- produces, on average, a gain in energy in the GeV/nucleon
region. Finally, spallations may destroy CRs, preferentially at low energies.

\vskip 0.1cm
Whatever the model retained for propagating antiprotons, it is very
important to understand the origin of uncertainties in the propagated
spectra. At a given energy $E$, spatial transport is sensitive to the
following parameters -- see Fig.~2 in \cite{Revue}~:
the diffusion coefficient normalization and slope in
$K(E) = \beta \, K_{0} \, {\cal R}^{- \delta}$ where
the rigidity ${\cal R} = p / Z$,
the halo height $L$,
the wind velocity $V_c$ perpendicular to the disk -- choosen to be
constant in our model --
and the Alfv\'enic speed $V_a$ of the scatterers -- the uncertainty due to
this latter parameter is less significant compared to the previous processes.
Only special combinations of these parameters can account for the measured B/C
ratio. The abundance of boron relative to carbon -- two typical elements which
are respectively from secondary and primary origin -- is a very good tracer of
the history of CRs propagation. In a previous study, a degeneracy between these
combinations was found \cite{Usine1}, leading to a wide uncertainty in the underlying
parameters of the model, although they gave the same B/C ratio. In a second study,
the same combinations were used to compute the secondary antiproton signal in the
same model \cite{Usine2}. The induced uncertainty for this secondary flux in the
region of interest -- a few hundreds of MeV -- was found to be small -- about 10\%.
This was expected, as all these species follow the same propagation paths, being
emitted and detected in the disk.

\vskip 0.1cm
The situation is quite different for primary exotic species, as most diffusion
paths start in the diffusive halo \cite{TailletMaurin, MaurinTaillet}. The previous
degeneracy is broken. The induced uncertainties on primary antiprotons are studied
in details in \cite{pbar_susy} and are found to be as large as a factor $\sim$ 100
for supersymmetric particles. We briefly recall here, on a physical basis, the
dependance of the uncertainty on each parameter, as it also applies to the present
study.
First, the halo height $L$ determines
i) the total number of sources inside the diffusive region and
ii) the effective radial range of diffusion, i.e. the distance that a CR can travel
from a source before escaping from the Galaxy. Cosmic rays coming from farther than
$L$ have an exponentially low probability to be detected on Earth. Notice that this
second point explains why the evaluated fluxes are not very sensitive to the shape
of the dark matter halo in the galactic center region -- see below in section~IV and
in \cite{pbar_susy}.
Second, the galactic wind wipes the particles away from the disk. It is well known
that the effect of $V_{c}$ is similar to that of $L$ when sources are located in
the disk -- see \cite{1978ApJ...222.1097J} and $L^{\star} \sim K(E)/L$ -- but this
is not true for sources in the halo. The two effects i) and ii) actually turn out to
be of greater magnitude for $L^{\star}$ than for $L$.

\vskip 0.1cm
It should be kept in mind that the parameters $L$, $V_{c}$ and $K_{0}$ are correlated.
In the subset of parameters giving the observed B/C ratio, low values of $K_{0}$ generally
correspond to low $L$ and large $V_{c}$ and thus low $L^{\star}$, so that the signal is
expected to decrease with decreasing $K_{0}$. Notice that the effect of $K_{0}$ is not
only through the correlation to $L$ and $V_{c}$~: the reader is referred to \cite{pbar_susy}
-- see in particular its section~III -- for an explicit analysis of all the effects.

\vskip 0.1cm
A conservative estimate -- based on the full range of B/C allowed propagation parameters
-- leads to variations of about two orders of magnitude of the primary antiproton flux.
Notice that this does not include the nuclear and particle physics uncertainties.
This range could be narrowed by using constraints coming from other species of cosmic rays.
Actually, using radioactive \cite{2002A&A...381..539D} or heavy \cite{celine} species
only yield a minor improvement. They enable to shrink the parameter space but leave
unchanged the values leading to the extremal fluxes. The final extreme and median parameters
which we have considered in this analysis are borrowed from \cite{pbar_susy}. They are
displayed in Tab.~\ref{table:prop}.

%
\begin{table}[h!]
\begin{center}
{\begin{tabular}{@{}|c|c|c|c|c|@{}}
\hline
{\rm case} &  $\delta$  & $K_0$                 & $L$   & $V_c$    \\
           &            & [${\rm kpc^{2}/Myr}$] & [kpc] & [km/sec] \\
\hline
\hline
{\rm max} &  0.46  & 0.0765 & 15 & 5    \\
{\rm med} &  0.70  & 0.0112 & 4  & 12   \\
{\rm min} &  0.85  & 0.0016 & 1  & 13.5 \\
\hline
\end{tabular}}
\caption{
Astrophysical parameters giving the maximal, median and minimal
LKP antiproton flux compatible wih B/C analysis.
\label{table:prop}}
\end{center}
\end{table}
%

\begin{figure}[h!]
\scalebox{0.80}
{\centerline{\includegraphics{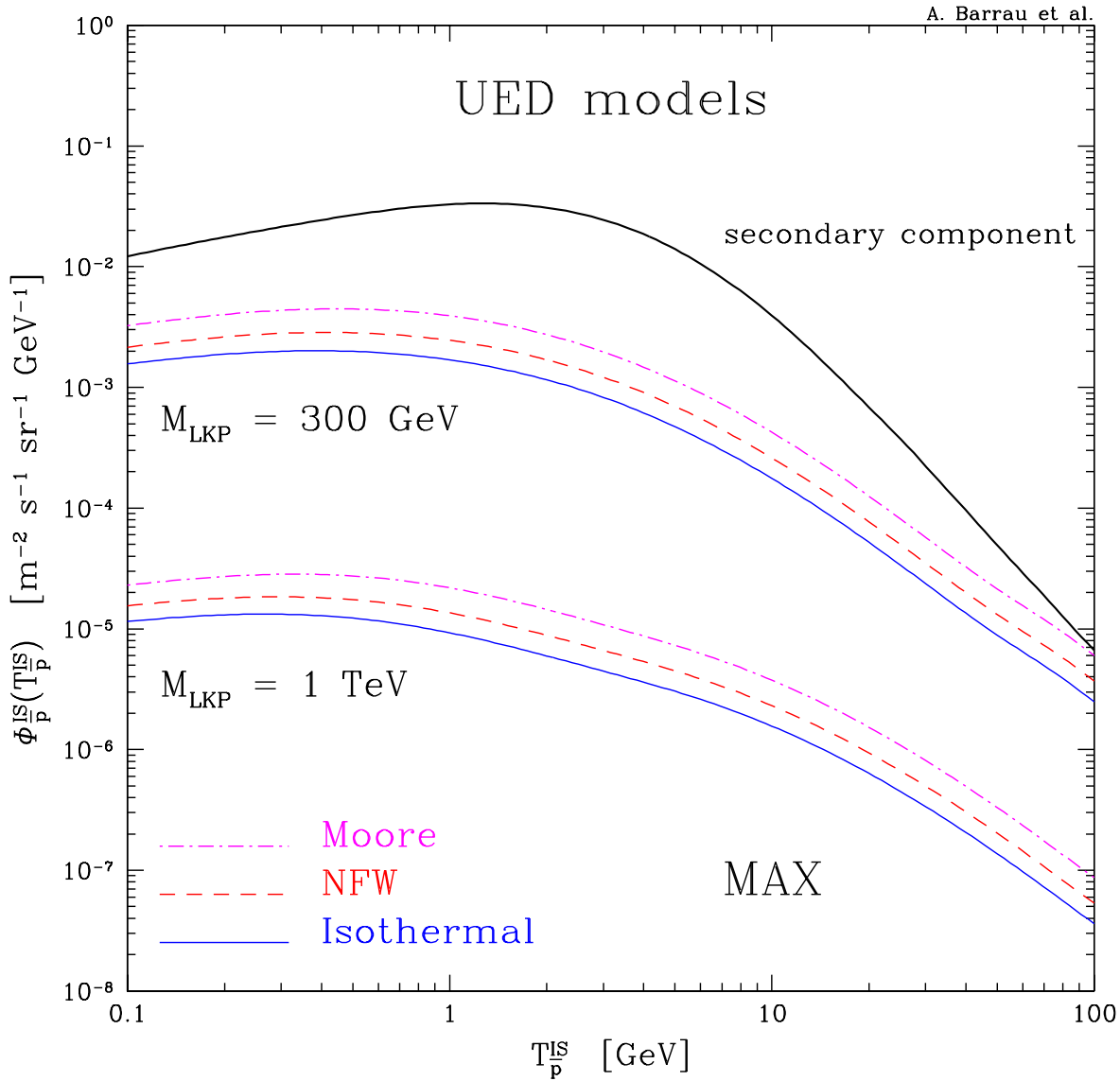}}}
\vskip -0.3cm
\caption{
The primary interstellar antiproton flux is featured as a function of the
antiproton kinetic energy $T_{\bar{p}}^{\rm IS}$ for a LKP mass of
300 GeV and 1 TeV. Maximal values for the diffusion parameters have been
assumed here -- see Tab.~\ref{table:prop}. The three different halo profiles
that have been selected in this calculation are described in Tab.~\ref{tab:indices}.
The IS secondary component is the solid black line that overcomes the primary
fluxes.}
\label{fig:UED_max}
\end{figure}

\begin{figure}[h!]
\scalebox{0.80}
{\centerline{\includegraphics{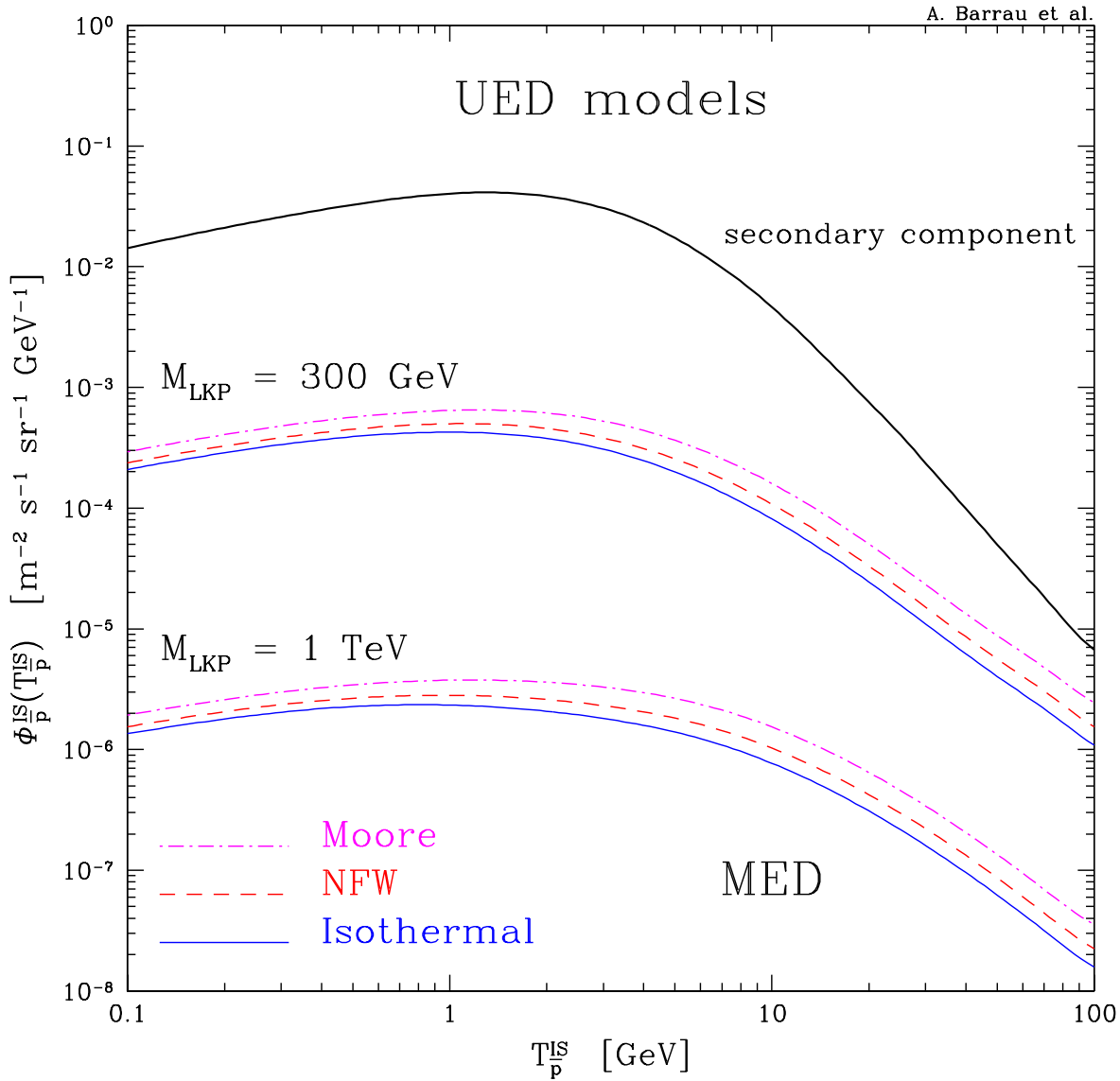}}}
\vskip -0.3cm
\caption{
The same as before but with median diffusion parameters. Primary fluxes
drop by one order of magnitude.}
\label{fig:UED_med}
\end{figure}

\begin{figure}[h!]
\scalebox{0.80}
{\centerline{\includegraphics{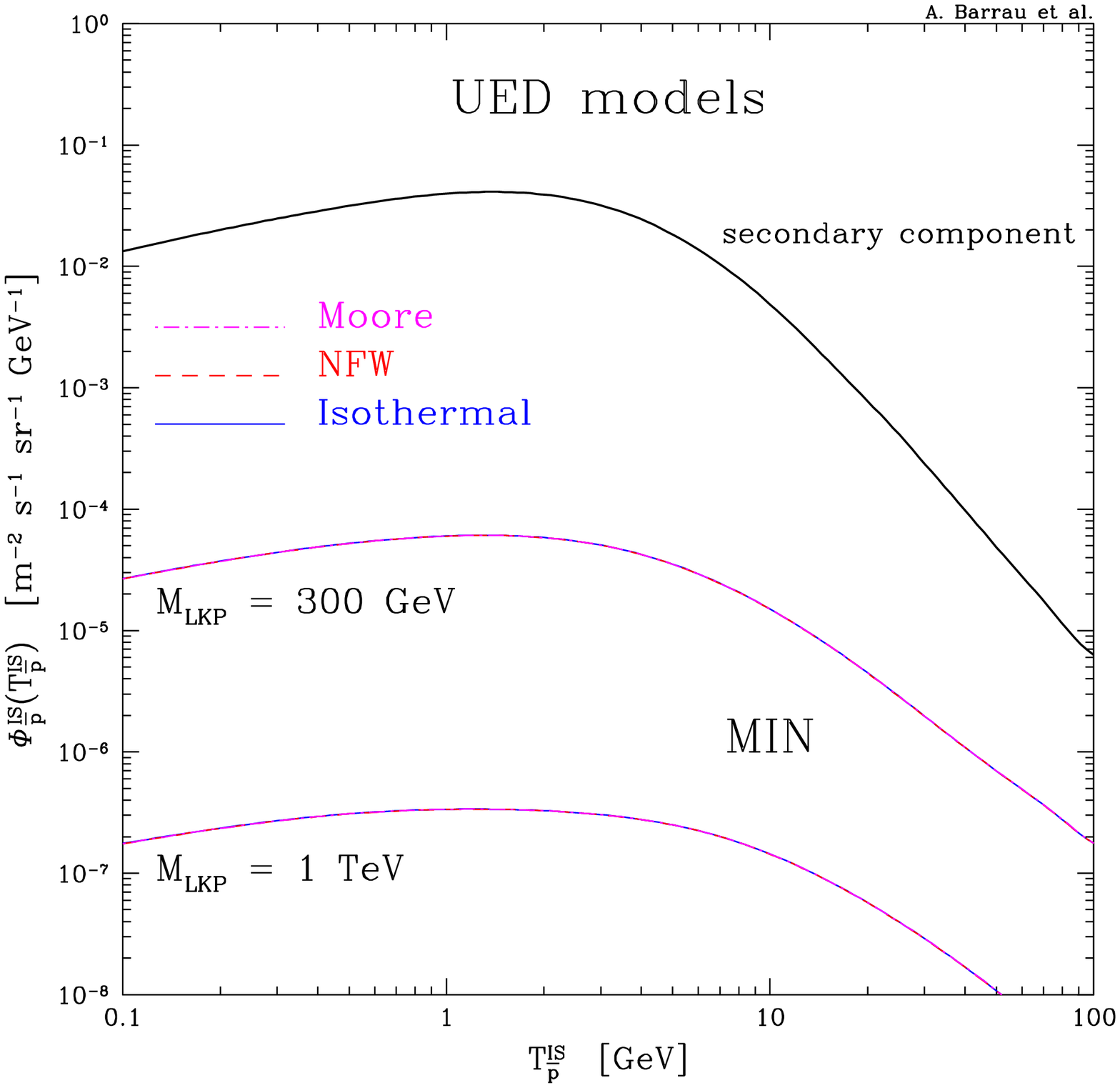}}}
\vskip -0.3cm
\caption{
The same as before but with minimum diffusion parameters. Fluxes have
decreased by two orders of magnitude with respect to the maximal case.
Cosmic rays no longer come from the galactic center. As a consequence,
the primary component is insensitive to the sharpness of the
central cusp and the three different halo profiles that have been
chosen in this calculation -- isothermal, NFW and Moore -- lead to
the same spectra.}
\label{fig:UED_min}
\end{figure}

\section{Flux and conclusions}

The case of UED models is featured in Fig.~\ref{fig:UED_max} to \ref{fig:UED_min}
where the LKP mass has been set equal to 300 GeV and to 1 TeV. The interstellar
antiproton yields are plotted as a function of interstellar kinetic energy
$T_{\bar{p}}^{\rm IS}$ for three different halo profiles. The canonical isothermal,
NFW and Moore models respectively correspond to the solid blue, dashed red and
dot-dashed magenta curves. The solid black line is the conventional secondary
component. We have somewhat improved the previous estimate \cite{Usine2} by
taking adiabatic losses into account. The maximum, median and minimum diffusion
configurations respectively correspond to Fig.~\ref{fig:UED_max}, \ref{fig:UED_med}
and \ref{fig:UED_min}. A few remarks are in order.

\vskip 0.1cm
\noindent
To commence, because the square of the LKP mass enters into the denominator
of the $B^{(1)}$ annihilation cross section -- see relation~(\ref{cs_UED})
-- the antiproton source term $q_{\bar{p}}^{\rm LKP}$ varies globally like
$M_{\rm LKP}^{-4}$. As a consequence, when the LKP mass is increased from
300 GeV to 1 TeV, the antiproton fluxes drop by a factor of
$(10 / 3)^{4} \sim 120$. This downward shift of the curves by two orders
of magnitude is clearly present in the figures.

\vskip 0.1cm
\noindent
Then, as already discussed in section~III, the particular choice for
the galactic cosmic ray diffusion parameters strongly affects the primary
yields whereas the secondary component varies very little. From the
maximal to the minimal configurations -- see Tab.~\ref{table:prop} --
primary antiproton fluxes decrease by two orders of magnitude. That
sensitivity combined with fairly similar shapes for the primary and
secondary energy spectra do not strengthen the case of the antiproton
signal as a clear signature for UED dark matter. Even in the most favorable
case of a 300 GeV $B^{(1)}$ boson and for maximal galactic diffusion
-- see Fig.~\ref{fig:UED_max} -- the secondary background overcomes
the primary LKP signal up to an antiproton kinetic energy of 100 GeV.
Notice however that above that energy and in the case of a Moore halo
profile, the signal may eventually become comparable to the background,
leading to an excess of antiprotons at high energy that has already been
noticed by \cite{torsten}. Unfortunately, that distinctive spectral
feature vanishes as soon as other configurations for the galactic cosmic
ray propagation are selected. In Fig.~\ref{fig:UED_min}, the antiproton
yield is $\sim$ 30 times smaller than the secondary flux for an antiproton
kinetic energy of 100 GeV.

\vskip 0.1cm
\noindent
Finally, the cusp at the Milky Way center does not affect much the
primary antiproton signal. Varying the DM halo profile from a mild
canonical isothermal distribution to the extreme case of a Moore
divergence results in an increase of the primary yields by at most
a factor $\sim$ 2 to 3 in the case of maximal galactic diffusion.
That increase is much less significant for the median diffusion
case and has disappeared in Fig.~\ref{fig:UED_min}. As is clear in
Tab.~\ref{table:prop}, the minimal diffusion configuration corresponds
to a thickness of the confinement layers of only 1 kpc associated with
a strong galactic convection wind that wipes away any particle
originating from the galactic central cusp.

\begin{figure}[t!]
\scalebox{0.80}
{\centerline{\includegraphics{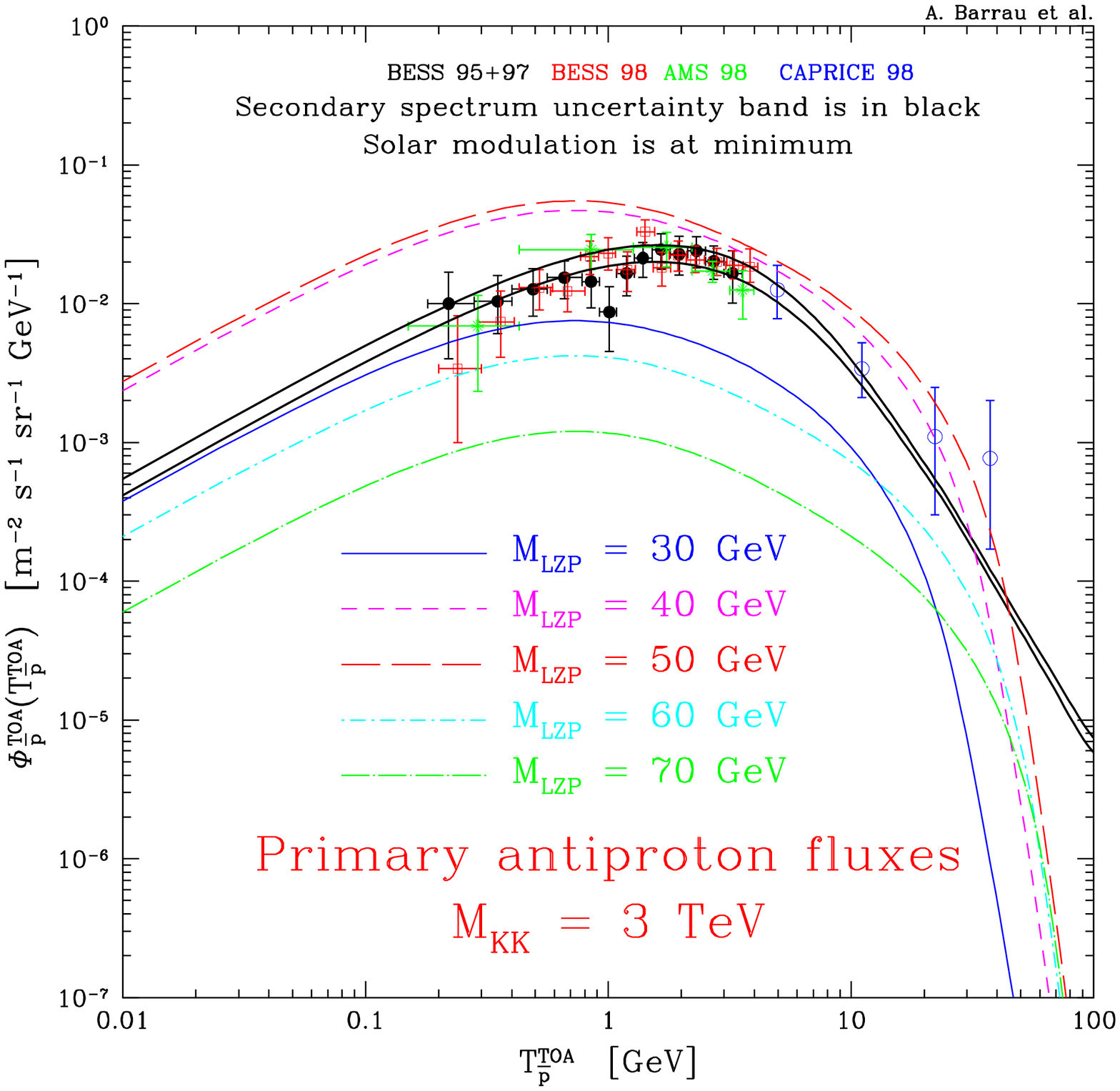}}}
\vskip -0.3cm
\caption{
The primary antiproton fluxes correspond to the warped geometry of
\cite{Agashe1,Agashe2}. The LZP mass has been varied from 30 to 70 GeV
with a Kaluza-Klein scale $M_{\rm KK}$ of 3 TeV. When the LZP mass is
close to $M_{\rm Z^{0}} / 2$, the annihilation becomes resonant and the
primary signal exceeds the conventional background of secondary antiprotons
that lies in the narrow  band within the two solid black lines.
Maximal diffusion parameters have been assumed for the LZP antiproton
spectra with a canonical isothermal DM distribution.}
\label{fig:LZP_mass}
\end{figure}

\begin{figure}[t!]
\scalebox{0.80}
{\centerline{\includegraphics{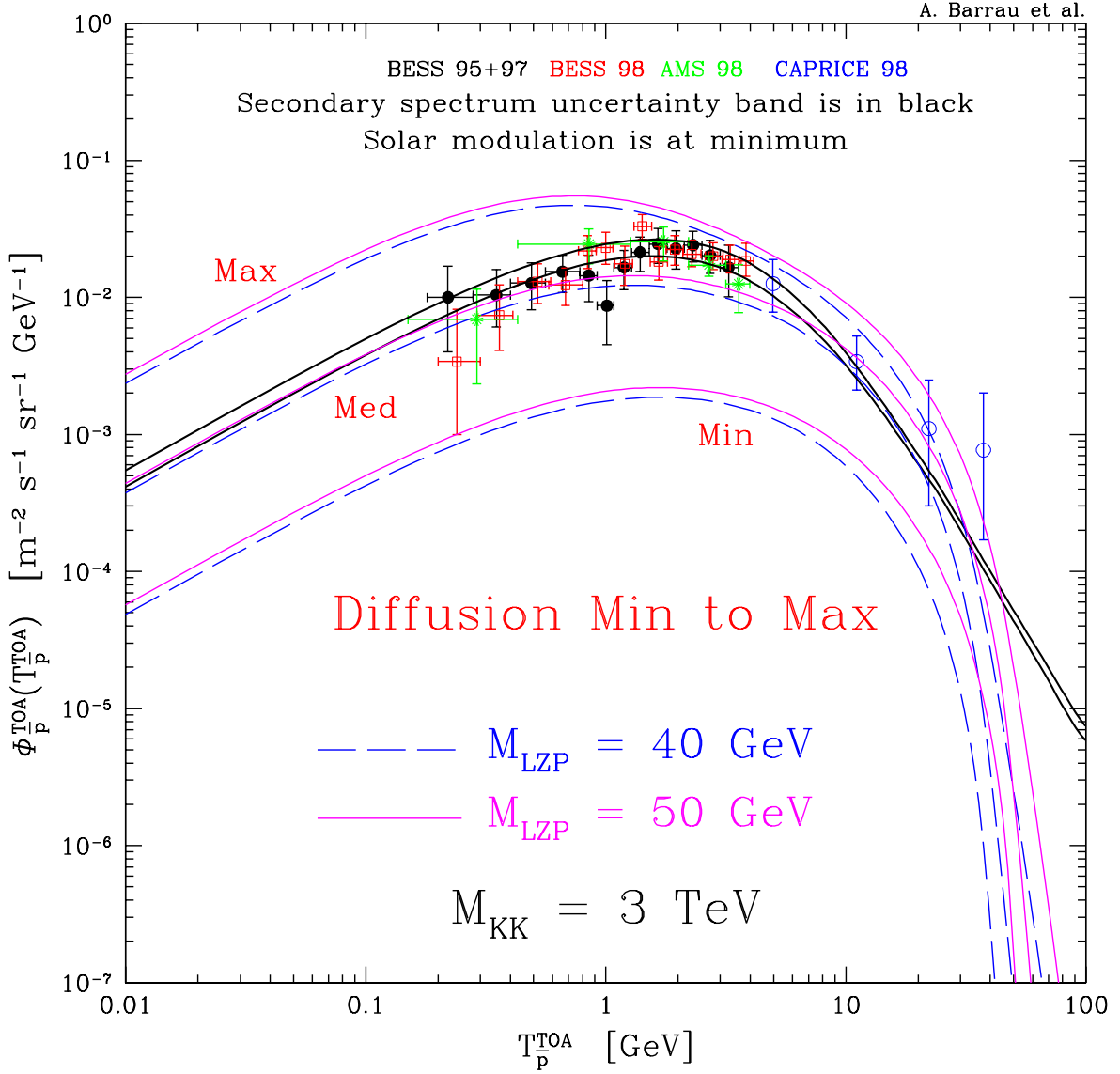}}}
\vskip -0.3cm
\caption{
When the diffusion parameters are varied over the entire domain that is
compatible with the B/C ratio, antiproton primary fluxes span two orders
of magnitude whilst the secondary component lies within a much narrower
band. The case of a resonant LZP has been featured here with
$M_{\rm LZP} = 40$ -- blue dashed -- and 50 GeV -- solid magenta.
A canonical isothermal DM distribution has been assumed. In the case of
minimal diffusion, the LZP signal is well below the background.}
\label{fig:LZP_diffusion}
\end{figure}

\begin{figure}[t!]
\scalebox{0.80}
{\centerline{\includegraphics{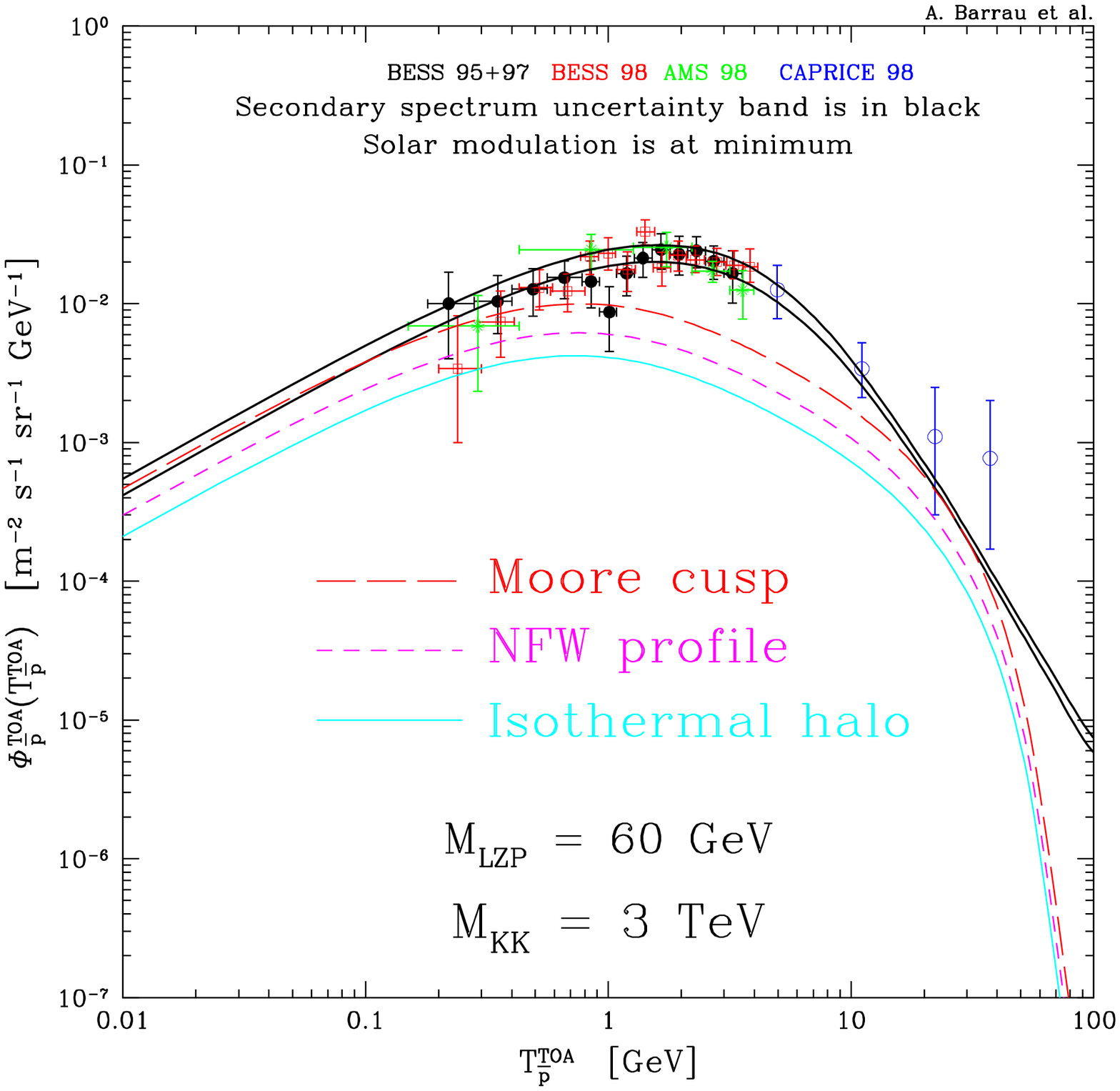}}}
\vskip -0.3cm
\caption{
The effect of the DM halo profile is presented in this figure where
the mass of the LZP has been set equal to $M_{\rm LZP} = 60$ GeV with
a Kaluza-Klein scale $M_{\rm KK}$ of 3 TeV. The more divergent and
concentrated the LZP distribution at the center of the Milky Way,
the larger the antiproton yield. That effect is particularly acute
in this plot where maximum diffusion parameters have been assumed.}
\label{fig:LZP_halo}
\end{figure}

\vskip 0.1cm
\noindent
Notice that our galactic diffusion code relies on the expansion
of the radial dependence of the cosmic ray abundances as a series
of the Bessel functions $J_{0}(\alpha_{i} \, r / R_{\rm gal})$
where $\alpha_{i}$ is the i-th zero of the function $J_{0}$ and
where $R_{\rm gal}$ is the radius of the propagation region.
Because taking properly into account a central divergence like
$r^{- 2 \gamma}$ with $\gamma = 1$ -- NFW -- or 1.3 -- Moore --
would necessitate an infinite number of such functions in the
above expansion and would lead to numerical instability, we have
renormalized the DM distribution in the vicinity of the Milky Way
center without modifying the absolute number of its annihilations.
More precisely, the actual DM density within a sphere of radius $r_{c}$
is given by
\be
{\displaystyle \frac{\rho(r)}{\rho_{c}}} \; = \; \left\{
{\displaystyle \frac{r_{c}}{r}} \right\}^{\gamma} \;\; ,
\label{profile_divergent}
\ee
where $\rho_{c} \equiv \rho(r_{c})$. The central cusp boosts
the LKP annihilations by a factor of
\be
\eta \; = \;
{\displaystyle \frac{3}{3 \, - \, 2 \, \gamma}}
\ee
with respect to the case of a uniform distribution with constant
density $\rho_{c}$. We have replaced the divergent
distribution~(\ref{profile_divergent}) by the milder profile
\be
\left\{
{\displaystyle \frac{\rho(r \leq r_{c})}{\rho_{c}}} \right\}^{2}
\; = \; 1 \; + \; \left\{
{\displaystyle \frac{2 \, \pi^{2}}{3}} \,
\left( \eta - 1 \right) \,
\sin_{c}^{2} \left( {\displaystyle \frac{\pi \, r}{r_{c}}} \right)
\right\} \;\; ,
\label{profile_convergent}
\ee
where $\sin_{c}(x) \equiv \sin(x) / x$. That renormalized density
leads to the same number of LKP annihilations as the actual cusp.
We have set $r_{c} = 500$ pc and our primary flux calculations
converge with only $N_{\rm Bes} = 300$ terms in the Bessel expansion.
A smaller value of $r_{c}$ would require a larger $N_{\rm Bes}$
and is not actually necessary insofar as the antiproton Green function
that connects the solar system to the galactic central region varies
smoothly over the latter \cite{TailletMaurin, MaurinTaillet}. In the minimal
case for cosmic ray propagation -- presented in Fig.~\ref{fig:UED_min}
-- it even vanishes.

\vskip 0.1cm
\noindent
Because of the uncertainties in the cosmic ray galactic propagation,
we conclude that the antiproton signal is not the best tool to observe
UED Kaluza-Klein species in the halo of the Milky Way. Direct detection
is not very promising either since observable rates at current instruments
are typically less than one event per year \cite{directLKP}. On the
contrary, since a pair of $B^{(1)}$ bosons may annihilate directly into
light fermions, the positron signal should exhibit a characteristic
spectral spike spreading toward low energies as a result of positron
energy losses during propagation \cite{cheng}. Notice however that
the positron annihilation signal arising in the case of an isothermal
halo needs to be amplified by a factor of $\sim$ 60 \cite{jonathan}
before being detectable by AMS-02 \cite{ams}. An enhancement by a factor
of $\sim$ 200 with respect to a pure NFW cusp is also necessary to reproduce
-- albeit below 0.8 TeV \cite{gammas1} -- the flat gamma ray spectrum which
the HESS collaboration has detected at the center of the Milky Way \cite{hess}.
The positron and gamma ray spectra are harder for Kaluza-Klein species
than for neutralinos. Those signals have therefore been advocated as
promising signatures.
A word of caution is in order at that stage. The positron signature
requires to be significantly enhanced in order to be detectable.
If we now assume a boost factor of $\sim$ a hundred as suggested by
recent numerical simulations \cite{mini_clump} that point toward the
presence of numerous mini-clumps in the DM galactic halo, primary
antiprotons should be copiously produced since even in the most
pessimistic diffusion scheme of Fig.~\ref{fig:UED_min}, the signal
exceeds the background above $T_{\bar{p}}^{\rm IS} \sim 40$ GeV
in the case of a 300 GeV $B^{(1)}$ boson.


\vskip 0.2cm
For the warped models of \cite{Agashe1,Agashe2}, the LZP may be much lighter
than the UED dark matter candidate. The primary antiproton flux -- at the
top of the atmosphere -- is plotted in Fig.~\ref{fig:LZP_mass} as a function
of antiproton kinetic energy $T_{\bar{p}}^{\rm TOA}$ for five different
values of the LZP mass. The most optimistic galactic diffusion scheme as well
as a canonical isothermal DM halo have been assumed for the primary signal.
Observations from various experiments performed during solar minimum
\cite{bess95_97,bess_98,ams_98,caprice_98} are well explained by the narrow
band within which the background of secondary antiprotons lies irrespective
of the galactic propagation conditions. The curves corresponding to
$M_{\rm LZP} = 40$ -- short dashed magenta -- and 50 GeV -- long dashed red --
exceed the background and should have already led to a detection would
our assumptions on galactic diffusion and halo profile be correct. For
$M_{\rm LZP} = M_{\rm Z^{0}} / 2$, the LZP annihilation is actually driven
by the $Z$-resonance and is significantly enhanced.

\vskip 0.1cm
\noindent
As in the previous discussion of the UED models, the LZP antiproton signal
sensitively depends on galactic cosmic ray propagation. In
Fig.~\ref{fig:LZP_diffusion}, the primary yields of a 40 and 50 GeV LZP
decrease by two orders of magnitude between the most optimistic and the most
pessimistic diffusion cases of Tab.~\ref{table:prop}. In the latter configuration,
the antiproton signal is now well below the background. The halo profile is also
a source of uncertainty as is clear in Fig.~\ref{fig:LZP_halo} where a 60 GeV LZP
is exhibited. The maximal galactic diffusion that has been assumed in that case
makes it possible for antiprotons from the central cusp to reach the solar circle
and to lift the degeneracy among the various DM distributions. Should the minimal
diffusion scheme be preferred, the three colored curves would be one and the same.

\begin{figure}[h!]
\scalebox{0.80}
{\centerline{\includegraphics{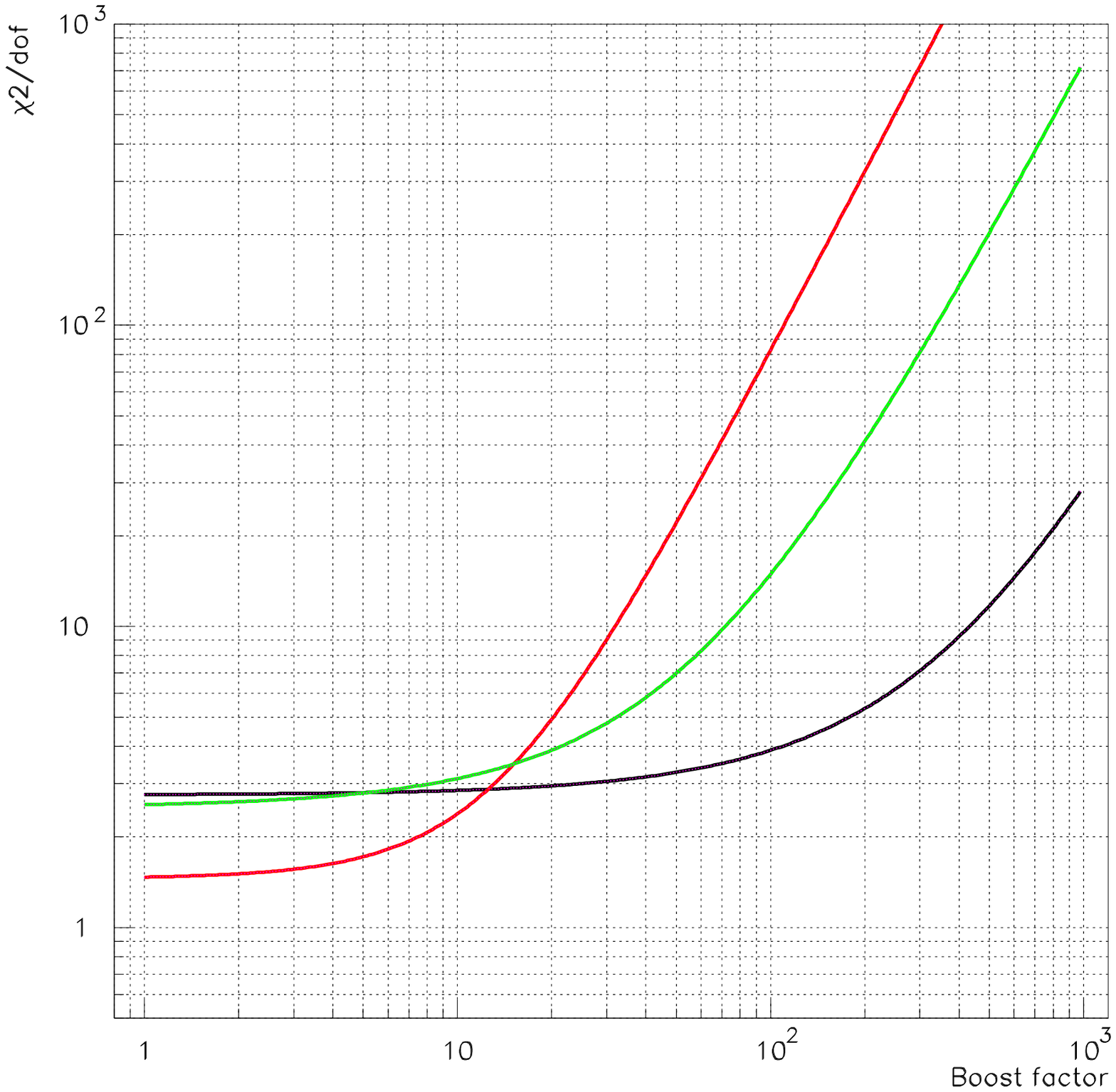}}}
\vskip -0.3cm
\caption{
The $\chi^{2}$ per degree of freedom is presented as a function of the boost
factor for $M_{\rm LZP} = 30$ GeV and $M_{\rm KK} = 3$ TeV. From top to bottom
-- at a boost factor of 100 -- the curves correspond to the maximum, median and
minimum diffusion parameters as defined in Tab.~\ref{table:prop}. In the left
part of the figure, the $\chi^{2}$ value is dominated by the secondary component
while in the right part, it is dominated by the primary component.}
\label{fig:chi2}
\end{figure}

%
\begin{table}[h!]
\begin{center}
\begin{tabular}{|p{3.0cm}|*{4}{c|}|}
\hline
LZP mass & Minimum diffusion & Median diffusion & Maximum diffusion\\
\hline
\hline
$M_{\rm LZP}$ = 30~GeV & $M_{\rm KK}$ = 3~TeV : 13.1 & $M_{\rm KK}$ = 3~TeV : 2.08     & $M_{\rm KK}$ = 3~TeV : excluded\\
                       & $M_{\rm KK}$ = 6~TeV : 214  & $M_{\rm KK}$ = 6~TeV : 33.10    & $M_{\rm KK}$ = 6~TeV : 12.9\\
\hline
$M_{\rm LZP}$ = 40~GeV & $M_{\rm KK}$ = 3~TeV : 2.09 & $M_{\rm KK}$ = 3~TeV : excluded & $M_{\rm KK}$ = 3~TeV : excluded\\
                       & $M_{\rm KK}$ = 6~TeV : 33.9 & $M_{\rm KK}$ = 6~TeV : 5.25     & $M_{\rm KK}$ = 6~TeV :  2.09\\
\hline
$M_{\rm LZP}$ = 50~GeV & $M_{\rm KK}$ = 3~TeV : 1.74 & $M_{\rm KK}$ = 3~TeV : excluded & $M_{\rm KK}$ = 3~TeV : excluded\\
                       & $M_{\rm KK}$ = 6~TeV : 27.5 & $M_{\rm KK}$ = 6~TeV : 4.27     & $M_{\rm KK}$ = 6~TeV : 2.70 \\
\hline
$M_{\rm LZP}$ = 60~GeV & $M_{\rm KK}$ = 3~TeV : 22.9 & $M_{\rm KK}$ = 3~TeV : 3.55     & $M_{\rm KK}$ = 3~TeV : 1.41\\
                       & $M_{\rm KK}$ = 6~TeV : 355  & $M_{\rm KK}$ = 6~TeV : 55.0     & $M_{\rm KK}$ = 6~TeV : 22.4 \\
\hline
$M_{\rm LZP}$ = 70~GeV & $M_{\rm KK}$ = 3~TeV : 79.4 & $M_{\rm KK}$ = 3~TeV : 12.3     & $M_{\rm KK}$ = 3~TeV : 5.01\\
                       & $M_{\rm KK}$ = 6~TeV : 1240 & $M_{\rm KK}$ = 6~TeV : 191      & $M_{\rm KK}$ = 6~TeV : 78.2\\
\hline
$M_{\rm LZP}$ = 80~GeV & $M_{\rm KK}$ = 3~TeV : 191  & $M_{\rm KK}$ = 3~TeV : 29.3     & $M_{\rm KK}$ = 3~TeV : 12.0\\
                       & $M_{\rm KK}$ = 6~TeV : 2930 & $M_{\rm KK}$ = 6~TeV : 464      & $M_{\rm KK}$ = 6~TeV : 185\\
\hline
\hline
\end{tabular}
\end{center}
\caption{
Boost factor above which the LZP model is excluded. The condition that
$\chi^{2} / {\rm d.o.f.} \; \geq \; 2 \times \chi^{2}_{\rm min} / {\rm d.o.f.}$
is required. Different galactic diffusion schemes have been considered.
An isothermal DM halo profile is assumed.}
\label{chi2_tab}
\end{table}
%

\vskip 0.1cm
\noindent
As featured in Fig.~\ref{fig:LZP_mass} to \ref{fig:LZP_halo}, the LZP
antiproton signal is in the vicinity of the secondary background and
therefore in the ballpark for detection. That is why we have explored
the effect of a clumpy DM distribution by taking into account an overall
boost factor in our estimates of the primary yields which we have compared
to observations. We have actually performed a $\chi^{2}$ test to assess
the compatibility between our theoretical predictions for both secondary
and primary components and the experimental data. All the available
measurements have been used
\cite{bess95_97,bess_98,bess_00,ams_98,caprice_98,caprice_94,imax,mass_91}
except the Buffington point which is known for being one order of magnitude
above all the others. Those experiments are either balloon borne -- IMAX, MASS,
CAPRICE and BESS -- or space borne as AMS. In addition to statistical effects,
they suffer from uncertainties associated with instrumental misreconstructions
-- {\it e.g.} from electrons -- and from the atmosphere component contamination
which has to be removed -- unless only antiprotons above the geomagnetic cutoff
are taken into account. Within the error bars, the IS fluxes inferred from those
experiments are now in reasonable agreement. A dramatic improvement is expected
in the forthcoming years with AMS-02 which will be implemented on the International
Space Station for 3 years starting in 2008~: both statistic and systematic
errors are expected to be reduced by several orders of magnitude for antiprotons
above 0.5~GeV.
To take into account the differences in solar activity between those observations,
the modulation has been applied in the force field scheme with three different
field values $\phi$~: 500 MV, 700 MV and 1000 MV, depending on the periods. The errors
used are the experimental statistical uncertainties. Fig.~\ref{fig:chi2} features
the $\chi^{2}$ per degree of freedom as a function of the boost factor for
$M_{\rm LZP} = 30$ Gev and a Kaluza-Klein scale $M_{\rm KK}$ of 3 TeV.
The red, green and black curves respectively correspond to the maximum, median
and minimum cosmic ray diffusion configurations.
It should be emphasized that the value of that $\chi^{2}$ must be taken with
care as it indicates that the uncertainties have been underestimated, making
any quantitative statistical conclusion impossible to reach. To give a crude
estimate of the rejection power of this study, we have decided that models
leading to a $\chi^{2} / {\rm d.o.f.}$ larger than twice its minimum value
-- for the considered parameters -- are excluded. In the case of a real
$\chi^{2}$ distribution, it would correspond roughly to the 99.9 \% confidence
level -- we use $\sim 50$ degrees of freedom.
In Tab.~\ref{chi2_tab}, the LZP mass has been varied from 30 to 80 GeV whereas
the Kaluza-Klein scale $M_{\rm KK}$ has been set equal to 3 and 6 TeV. The DM
annihilation boost factor above which the primary antiproton signal is too
strong to be compatible with the observations is displayed for each configuration.
If a boost factor of $\sim$ one hundred is assumed, all the configurations with
$M_{\rm KK}$ = 3 TeV are excluded whereas the LZP antiproton signal is potentially
detectable for larger Kaluza-Klein scales.

\vskip 0.1cm
\noindent
Notice finally that direct detection experiments already exclude almost entirely
a Kaluza-Klein scale $M_{\rm KK}$ of 3 TeV whereas larger values are allowed
\cite{Agashe1}. The LZP may also directly annihilate into light fermions and can
produce the same kind of distortion in the positron spectrum as UED dark matter
species. The HEAT excess \cite{heat_1,heat_2} is actually well reproduced by a 40
or 50 GeV LZP if the boost factor is respectively set equal to $\sim$ 40 and 30
\cite{LZPindirect}.
Antiproton calculations suffer from large uncertainties as regards the galactic
cosmic ray propagation. The difficulty to reach a conclusion as regards the
detectability of a primary antiproton signal has been illustrated in this article.
We would like to stress that the same kind of ambiguities should also affect secondary
and primary positrons with a magnitude that is yet to be determined.

\newpage
\noindent
Acknowledgements : P.S. would like to thank the french programme national de cosmologie
PNC and the groupement de recherche on ph\'enom\`enes cosmiques de haute \'energie PCHE
for their financial support.


\begin{thebibliography}{57}
\expandafter\ifx\csname natexlab\endcsname\relax\def\natexlab#1{#1}\fi
\expandafter\ifx\csname bibnamefont\endcsname\relax
  \def\bibnamefont#1{#1}\fi
\expandafter\ifx\csname bibfnamefont\endcsname\relax
  \def\bibfnamefont#1{#1}\fi
\expandafter\ifx\csname citenamefont\endcsname\relax
  \def\citenamefont#1{#1}\fi
\expandafter\ifx\csname url\endcsname\relax
  \def\url#1{\texttt{#1}}\fi
\expandafter\ifx\csname urlprefix\endcsname\relax\def\urlprefix{URL }\fi
\providecommand{\bibinfo}[2]{#2}
\providecommand{\eprint}[2][]{\url{#2}}

\bibitem[{\citenamefont{{Freedman} and {Turner}}(2003)}]{turner}
\bibinfo{author}{\bibfnamefont{W.~L.} \bibnamefont{{Freedman}}}
  \bibnamefont{and} \bibinfo{author}{\bibfnamefont{M.~S.}
  \bibnamefont{{Turner}}}, \bibinfo{journal}{Rev.\ Mod.\ Phys.}
  \textbf{\bibinfo{volume}{75}}, \bibinfo{pages}{1433} (\bibinfo{year}{2003}).

\bibitem[{\citenamefont{\mbox{D.~N. Spergel {\it et al.}}}(2003)}]{wmap}
\bibinfo{author}{\bibnamefont{\mbox{D.~N. Spergel {\it et al.}}}},
  \bibinfo{journal}{Astrophys.\ J.\ Suppl.} \textbf{\bibinfo{volume}{148}},
  \bibinfo{pages}{175} (\bibinfo{year}{2003}).

\bibitem[{\citenamefont{{Appelquist} et~al.}(2001)\citenamefont{{Appelquist},
  {Cheng}, and {Dobrescu}}}]{appel}
\bibinfo{author}{\bibfnamefont{T.}~\bibnamefont{{Appelquist}}},
  \bibinfo{author}{\bibfnamefont{H.~C.} \bibnamefont{{Cheng}}},
  \bibnamefont{and} \bibinfo{author}{\bibfnamefont{B.~A.}
  \bibnamefont{{Dobrescu}}}, \bibinfo{journal}{Phys.\ Rev.\ D}
  \textbf{\bibinfo{volume}{64}}, \bibinfo{pages}{035002}
  (\bibinfo{year}{2001}).

\bibitem[{\citenamefont{{Randall} and {Sundrum}}(1999)}]{rs}
\bibinfo{author}{\bibfnamefont{L.}~\bibnamefont{{Randall}}} \bibnamefont{and}
  \bibinfo{author}{\bibfnamefont{R.}~\bibnamefont{{Sundrum}}},
  \bibinfo{journal}{Phys.\ Rev.\ Lett.} \textbf{\bibinfo{volume}{83}},
  \bibinfo{pages}{3370} (\bibinfo{year}{1999}).

\bibitem[{\citenamefont{{Arkani-Hamed}
  et~al.}(1998)\citenamefont{{Arkani-Hamed}, {Dimopoulos}, and {Dvali}}}]{led}
\bibinfo{author}{\bibfnamefont{N.}~\bibnamefont{{Arkani-Hamed}}},
  \bibinfo{author}{\bibfnamefont{S.}~\bibnamefont{{Dimopoulos}}},
  \bibnamefont{and} \bibinfo{author}{\bibfnamefont{G.~R.}
  \bibnamefont{{Dvali}}}, \bibinfo{journal}{Phys.\ Lett.\ B}
  \textbf{\bibinfo{volume}{429}}, \bibinfo{pages}{263} (\bibinfo{year}{1998}).

\bibitem[{\citenamefont{{Cheng}
  et~al.}(2002{\natexlab{a}})\citenamefont{{Cheng}, {Matchev}, and
  {Schmaltz}}}]{cheng2}
\bibinfo{author}{\bibfnamefont{H.~C.} \bibnamefont{{Cheng}}},
  \bibinfo{author}{\bibfnamefont{K.~T.} \bibnamefont{{Matchev}}},
  \bibnamefont{and}
  \bibinfo{author}{\bibfnamefont{M.}~\bibnamefont{{Schmaltz}}},
  \bibinfo{journal}{Phys.\ Rev.\ D} \textbf{\bibinfo{volume}{66}},
  \bibinfo{pages}{036005} (\bibinfo{year}{2002}{\natexlab{a}}).

\bibitem[{\citenamefont{{Servant} and {Tait}}(2002{\natexlab{a}})}]{servant1}
\bibinfo{author}{\bibfnamefont{G.}~\bibnamefont{{Servant}}} \bibnamefont{and}
  \bibinfo{author}{\bibfnamefont{T.~M.~P.} \bibnamefont{{Tait}}},
  \bibinfo{journal}{Nucl.\ Phys.\ B} \textbf{\bibinfo{volume}{650}},
  \bibinfo{pages}{391} (\bibinfo{year}{2002}{\natexlab{a}}).

\bibitem[{\citenamefont{{Kakizaki} et~al.}(2005)\citenamefont{{Kakizaki},
  {Matsumoto}, {Sato}, and {Senami}}}]{relic2}
\bibinfo{author}{\bibfnamefont{M.}~\bibnamefont{{Kakizaki}}},
  \bibinfo{author}{\bibfnamefont{S.}~\bibnamefont{{Matsumoto}}},
  \bibinfo{author}{\bibfnamefont{Y.}~\bibnamefont{{Sato}}}, \bibnamefont{and}
  \bibinfo{author}{\bibfnamefont{M.}~\bibnamefont{{Senami}}},
  \textbf{\bibinfo{volume}{arXiv:hep-ph/0502059}} (\bibinfo{year}{2005}).

\bibitem[{\citenamefont{{Servant} and {Tait}}(2002{\natexlab{b}})}]{directLKP}
\bibinfo{author}{\bibfnamefont{G.}~\bibnamefont{{Servant}}} \bibnamefont{and}
  \bibinfo{author}{\bibfnamefont{T.~M.~P.} \bibnamefont{{Tait}}},
  \bibinfo{journal}{New J.\ Phys.} \textbf{\bibinfo{volume}{4}},
  \bibinfo{pages}{99} (\bibinfo{year}{2002}{\natexlab{b}}).

\bibitem[{\citenamefont{{Cheng}
  et~al.}(2002{\natexlab{b}})\citenamefont{{Cheng}, {Feng}, and
  {Matchev}}}]{cheng}
\bibinfo{author}{\bibfnamefont{H.-C.} \bibnamefont{{Cheng}}},
  \bibinfo{author}{\bibfnamefont{J.~L.} \bibnamefont{{Feng}}},
  \bibnamefont{and} \bibinfo{author}{\bibfnamefont{K.~T.}
  \bibnamefont{{Matchev}}}, \bibinfo{journal}{Phys.\ Rev.\ Lett.}
  \textbf{\bibinfo{volume}{89}}, \bibinfo{pages}{211301}
  (\bibinfo{year}{2002}{\natexlab{b}}).

\bibitem[{\citenamefont{{Bertone} et~al.}(2003)\citenamefont{{Bertone},
  {Servant}, and {Sigl}}}]{servant2}
\bibinfo{author}{\bibfnamefont{G.}~\bibnamefont{{Bertone}}},
  \bibinfo{author}{\bibfnamefont{G.}~\bibnamefont{{Servant}}},
  \bibnamefont{and} \bibinfo{author}{\bibfnamefont{G.}~\bibnamefont{{Sigl}}},
  \bibinfo{journal}{Phys.\ Rev.\ D} \textbf{\bibinfo{volume}{68}},
  \bibinfo{pages}{044008} (\bibinfo{year}{2003}).

\bibitem[{\citenamefont{{Bergstrom}
  et~al.}(2005{\natexlab{a}})\citenamefont{{Bergstrom}, {Bringmann},
  {Eriksson}, and {Gustafsson}}}]{gammas1}
\bibinfo{author}{\bibfnamefont{L.}~\bibnamefont{{Bergstrom}}},
  \bibinfo{author}{\bibfnamefont{T.}~\bibnamefont{{Bringmann}}},
  \bibinfo{author}{\bibfnamefont{M.}~\bibnamefont{{Eriksson}}},
  \bibnamefont{and}
  \bibinfo{author}{\bibfnamefont{M.}~\bibnamefont{{Gustafsson}}},
  \bibinfo{journal}{Phys.\ Rev.\ Lett.} \textbf{\bibinfo{volume}{94}},
  \bibinfo{pages}{131301} (\bibinfo{year}{2005}{\natexlab{a}}).

\bibitem[{\citenamefont{{Baltz} and {Hooper}}(2004)}]{gammas2}
\bibinfo{author}{\bibfnamefont{E.~A.} \bibnamefont{{Baltz}}} \bibnamefont{and}
  \bibinfo{author}{\bibfnamefont{D.}~\bibnamefont{{Hooper}}},
  \textbf{\bibinfo{volume}{arXiv:astro-ph/0411053}} (\bibinfo{year}{2004}).

\bibitem[{\citenamefont{{Bergstrom}
  et~al.}(2005{\natexlab{b}})\citenamefont{{Bergstrom}, {Bringmann},
  {Eriksson}, and {Gustafsson}}}]{gammas3}
\bibinfo{author}{\bibfnamefont{L.}~\bibnamefont{{Bergstrom}}},
  \bibinfo{author}{\bibfnamefont{T.}~\bibnamefont{{Bringmann}}},
  \bibinfo{author}{\bibfnamefont{M.}~\bibnamefont{{Eriksson}}},
  \bibnamefont{and}
  \bibinfo{author}{\bibfnamefont{M.}~\bibnamefont{{Gustafsson}}},
  \bibinfo{journal}{JCAP} \textbf{\bibinfo{volume}{04}}, \bibinfo{pages}{004}
  (\bibinfo{year}{2005}{\natexlab{b}}).

\bibitem[{\citenamefont{{Hooper} and {Kribs}}(2004)}]{hooper2}
\bibinfo{author}{\bibfnamefont{D.}~\bibnamefont{{Hooper}}} \bibnamefont{and}
  \bibinfo{author}{\bibfnamefont{G.~D.} \bibnamefont{{Kribs}}},
  \bibinfo{journal}{Phys.\ Rev.\ D} \textbf{\bibinfo{volume}{70}},
  \bibinfo{pages}{115004} (\bibinfo{year}{2004}).

\bibitem[{\citenamefont{{Hooper} and {Kribs}}(2003)}]{hooper1}
\bibinfo{author}{\bibfnamefont{D.}~\bibnamefont{{Hooper}}} \bibnamefont{and}
  \bibinfo{author}{\bibfnamefont{G.~D.} \bibnamefont{{Kribs}}},
  \bibinfo{journal}{Phys.\ Rev.\ D} \textbf{\bibinfo{volume}{67}},
  \bibinfo{pages}{055003} (\bibinfo{year}{2003}).

\bibitem[{\citenamefont{{Kolb} et~al.}(2003)\citenamefont{{Kolb}, {Servant},
  and {Tait}}}]{KST}
\bibinfo{author}{\bibfnamefont{E.~W.} \bibnamefont{{Kolb}}},
  \bibinfo{author}{\bibfnamefont{G.}~\bibnamefont{{Servant}}},
  \bibnamefont{and} \bibinfo{author}{\bibfnamefont{T.~M.~P.}
  \bibnamefont{{Tait}}}, \bibinfo{journal}{JCAP} \textbf{\bibinfo{volume}{07}},
  \bibinfo{pages}{008} (\bibinfo{year}{2003}).

\bibitem[{\citenamefont{{Agashe} and {Servant}}(2004)}]{Agashe1}
\bibinfo{author}{\bibfnamefont{K.}~\bibnamefont{{Agashe}}} \bibnamefont{and}
  \bibinfo{author}{\bibfnamefont{G.}~\bibnamefont{{Servant}}},
  \bibinfo{journal}{Phys.\ Rev.\ Lett.} \textbf{\bibinfo{volume}{93}},
  \bibinfo{pages}{231805} (\bibinfo{year}{2004}).

\bibitem[{\citenamefont{{Agashe} and {Servant}}(2005)}]{Agashe2}
\bibinfo{author}{\bibfnamefont{K.}~\bibnamefont{{Agashe}}} \bibnamefont{and}
  \bibinfo{author}{\bibfnamefont{G.}~\bibnamefont{{Servant}}},
  \bibinfo{journal}{JCAP} \textbf{\bibinfo{volume}{02}}, \bibinfo{pages}{002}
  (\bibinfo{year}{2005}).

\bibitem[{\citenamefont{{Hooper} and {Servant}}(2005)}]{LZPindirect}
\bibinfo{author}{\bibfnamefont{D.}~\bibnamefont{{Hooper}}} \bibnamefont{and}
  \bibinfo{author}{\bibfnamefont{G.}~\bibnamefont{{Servant}}},
  \textbf{\bibinfo{volume}{arXiv:hep-ph/0502247}} (\bibinfo{year}{2005}).

\bibitem[{\citenamefont{{Donato} et~al.}(2001)\citenamefont{{Donato}, {Maurin},
  {Salati}, {Barrau}, {Boudoul}, and {Taillet}}}]{Usine2}
\bibinfo{author}{\bibfnamefont{F.}~\bibnamefont{{Donato}}},
  \bibinfo{author}{\bibfnamefont{D.}~\bibnamefont{{Maurin}}},
  \bibinfo{author}{\bibfnamefont{P.}~\bibnamefont{{Salati}}},
  \bibinfo{author}{\bibfnamefont{A.}~\bibnamefont{{Barrau}}},
  \bibinfo{author}{\bibfnamefont{G.}~\bibnamefont{{Boudoul}}},
  \bibnamefont{and}
  \bibinfo{author}{\bibfnamefont{R.}~\bibnamefont{{Taillet}}},
  \bibinfo{journal}{Astrophys.\ J.} \textbf{\bibinfo{volume}{563}},
  \bibinfo{pages}{172} (\bibinfo{year}{2001}).

\bibitem[{\citenamefont{\mbox{IMAX Collaboration -- J.~W. Mitchell {\it et
  al.}}}(1996)}]{imax}
\bibinfo{author}{\bibnamefont{\mbox{IMAX Collaboration -- J.~W. Mitchell {\it
  et al.}}}}, \bibinfo{journal}{Phys.\ Rev.\ Lett.}
  \textbf{\bibinfo{volume}{76}}, \bibinfo{pages}{3057} (\bibinfo{year}{1996}).

\bibitem[{\citenamefont{\mbox{MASS Collaboration -- G. Basini {\it et
  al.}}}(1999)}]{mass_91}
\bibinfo{author}{\bibnamefont{\mbox{MASS Collaboration -- G. Basini {\it et
  al.}}}}, \bibinfo{journal}{Proceedings of the 26th ICRC}
  (\bibinfo{year}{1999}).

\bibitem[{\citenamefont{\mbox{CAPRICE Collaboration -- M. Boezio {\it et
  al.}}}(2001)}]{caprice_98}
\bibinfo{author}{\bibnamefont{\mbox{CAPRICE Collaboration -- M. Boezio {\it et
  al.}}}}, \bibinfo{journal}{Astrophys.\ J.} \textbf{\bibinfo{volume}{561}},
  \bibinfo{pages}{787} (\bibinfo{year}{2001}).

\bibitem[{\citenamefont{\mbox{CAPRICE Collaboration -- M. Boezio {\it et
  al.}}}(1997)}]{caprice_94}
\bibinfo{author}{\bibnamefont{\mbox{CAPRICE Collaboration -- M. Boezio {\it et
  al.}}}}, \bibinfo{journal}{Astrophys.\ J.} \textbf{\bibinfo{volume}{487}},
  \bibinfo{pages}{415} (\bibinfo{year}{1997}).

\bibitem[{\citenamefont{\mbox{BESS Collaboration -- S. Orito {\it et
  al.}}}(2000)}]{bess95_97}
\bibinfo{author}{\bibnamefont{\mbox{BESS Collaboration -- S. Orito {\it et
  al.}}}}, \bibinfo{journal}{Phys.\ Rev.\ Lett.} \textbf{\bibinfo{volume}{84}},
  \bibinfo{pages}{1078} (\bibinfo{year}{2000}).

\bibitem[{\citenamefont{\mbox{BESS Collaboration -- T. Maeno {\it et
  al.}}}(2001)}]{bess_98}
\bibinfo{author}{\bibnamefont{\mbox{BESS Collaboration -- T. Maeno {\it et
  al.}}}}, \bibinfo{journal}{Astropart.\ Phys.} \textbf{\bibinfo{volume}{16}},
  \bibinfo{pages}{121} (\bibinfo{year}{2001}).

\bibitem[{\citenamefont{\mbox{BESS Collaboration -- Y. Asoaka {\it et
  al.}}}(2002)}]{bess_00}
\bibinfo{author}{\bibnamefont{\mbox{BESS Collaboration -- Y. Asoaka {\it et
  al.}}}}, \bibinfo{journal}{Phys.\ Rev.\ Lett.} \textbf{\bibinfo{volume}{88}},
  \bibinfo{pages}{051101} (\bibinfo{year}{2002}).

\bibitem[{\citenamefont{\mbox{AMS Collaboration}}(2005)}]{ams}
\bibinfo{author}{\bibnamefont{\mbox{AMS Collaboration}}},
  \bibinfo{journal}{http://pcamss0.cern.ch/mm.html}  (\bibinfo{year}{2005}).

\bibitem[{\citenamefont{\mbox{AMS Collaboration -- M. Aguilar {\it et
  al.}}}(2002)}]{ams_98}
\bibinfo{author}{\bibnamefont{\mbox{AMS Collaboration -- M. Aguilar {\it et
  al.}}}}, \bibinfo{journal}{Phys.\ Rep.} \textbf{\bibinfo{volume}{366}},
  \bibinfo{pages}{331} (\bibinfo{year}{2002}).

\bibitem[{\citenamefont{{Tjostrand}}(1994)}]{tj}
\bibinfo{author}{\bibfnamefont{T.}~\bibnamefont{{Tjostrand}}},
  \bibinfo{journal}{Comput.\ Phys.\ Commun.} \textbf{\bibinfo{volume}{82}},
  \bibinfo{pages}{74} (\bibinfo{year}{1994}).

\bibitem[{\citenamefont{{Navarro} et~al.}(1996)\citenamefont{{Navarro},
  {Frenk}, and {White}}}]{nfw}
\bibinfo{author}{\bibfnamefont{J.~F.} \bibnamefont{{Navarro}}},
  \bibinfo{author}{\bibfnamefont{C.~S.} \bibnamefont{{Frenk}}},
  \bibnamefont{and} \bibinfo{author}{\bibfnamefont{S.~D.~M.}
  \bibnamefont{{White}}}, \bibinfo{journal}{Astrophys.\ J.}
  \textbf{\bibinfo{volume}{462}}, \bibinfo{pages}{563} (\bibinfo{year}{1996}).

\bibitem[{\citenamefont{{Fukushige} and {Makino}}(1997)}]{fukushige}
\bibinfo{author}{\bibfnamefont{T.}~\bibnamefont{{Fukushige}}} \bibnamefont{and}
  \bibinfo{author}{\bibfnamefont{J.}~\bibnamefont{{Makino}}},
  \bibinfo{journal}{Astrophys.\ J.} \textbf{\bibinfo{volume}{477}},
  \bibinfo{pages}{L9} (\bibinfo{year}{1997}).

\bibitem[{\citenamefont{\mbox{B. Moore {\it et al.}}}(1999)}]{moore}
\bibinfo{author}{\bibnamefont{\mbox{B. Moore {\it et al.}}}},
  \bibinfo{journal}{Mon.\ Not.\ Roy.\ Astron.\ Soc.}
  \textbf{\bibinfo{volume}{310}}, \bibinfo{pages}{1147} (\bibinfo{year}{1999}).

\bibitem[{\citenamefont{{Diemand} et~al.}(2004)\citenamefont{{Diemand},
  {Moore}, and {Stadel}}}]{2004MNRAS.353..624D}
\bibinfo{author}{\bibfnamefont{J.}~\bibnamefont{{Diemand}}},
  \bibinfo{author}{\bibfnamefont{B.}~\bibnamefont{{Moore}}}, \bibnamefont{and}
  \bibinfo{author}{\bibfnamefont{J.}~\bibnamefont{{Stadel}}},
  \bibinfo{journal}{Mon.\ Not.\ Roy.\ Astron.\ Soc.}
  \textbf{\bibinfo{volume}{353}}, \bibinfo{pages}{624} (\bibinfo{year}{2004}).

\bibitem[{\citenamefont{{Navarro} et~al.}(2004)\citenamefont{{Navarro},
  {Hayashi}, {Power}, {Jenkins}, {Frenk}, {White}, {Springel}, {Stadel}, and
  {Quinn}}}]{2004MNRAS.349.1039N}
\bibinfo{author}{\bibfnamefont{J.~F.} \bibnamefont{{Navarro}}},
  \bibinfo{author}{\bibfnamefont{E.}~\bibnamefont{{Hayashi}}},
  \bibinfo{author}{\bibfnamefont{C.}~\bibnamefont{{Power}}},
  \bibinfo{author}{\bibfnamefont{A.~R.} \bibnamefont{{Jenkins}}},
  \bibinfo{author}{\bibfnamefont{C.~S.} \bibnamefont{{Frenk}}},
  \bibinfo{author}{\bibfnamefont{S.~D.~M.} \bibnamefont{{White}}},
  \bibinfo{author}{\bibfnamefont{V.}~\bibnamefont{{Springel}}},
  \bibinfo{author}{\bibfnamefont{J.}~\bibnamefont{{Stadel}}}, \bibnamefont{and}
  \bibinfo{author}{\bibfnamefont{T.~R.} \bibnamefont{{Quinn}}},
  \bibinfo{journal}{Mon.\ Not.\ Roy.\ Astron.\ Soc.}
  \textbf{\bibinfo{volume}{349}}, \bibinfo{pages}{1039} (\bibinfo{year}{2004}).

\bibitem[{\citenamefont{{Borriello} and {Salucci}}(2001)}]{bs}
\bibinfo{author}{\bibfnamefont{A.}~\bibnamefont{{Borriello}}} \bibnamefont{and}
  \bibinfo{author}{\bibfnamefont{P.}~\bibnamefont{{Salucci}}},
  \bibinfo{journal}{Mon.\ Not.\ Roy.\ Astron.\ Soc.}
  \textbf{\bibinfo{volume}{323}}, \bibinfo{pages}{285} (\bibinfo{year}{2001}).

\bibitem[{\citenamefont{{de Blok} and {Bosma}}(2002)}]{deblok}
\bibinfo{author}{\bibfnamefont{W.~J.~G.} \bibnamefont{{de Blok}}}
  \bibnamefont{and} \bibinfo{author}{\bibfnamefont{A.}~\bibnamefont{{Bosma}}},
  \bibinfo{journal}{Astron.\ Astroph.} \textbf{\bibinfo{volume}{385}},
  \bibinfo{pages}{816} (\bibinfo{year}{2002}).

\bibitem[{\citenamefont{\mbox{R.~A. Swaters {\it et al.}}}(2003)}]{swaters}
\bibinfo{author}{\bibnamefont{\mbox{R.~A. Swaters {\it et al.}}}},
  \bibinfo{journal}{Astrophys.\ J.} \textbf{\bibinfo{volume}{583}},
  \bibinfo{pages}{732} (\bibinfo{year}{2003}).

\bibitem[{\citenamefont{{Weldrake} et~al.}(2003)\citenamefont{{Weldrake}, {de
  Blok}, and {Walter}}}]{weldrake}
\bibinfo{author}{\bibfnamefont{D.~T.~F.} \bibnamefont{{Weldrake}}},
  \bibinfo{author}{\bibfnamefont{W.~J.~G.} \bibnamefont{{de Blok}}},
  \bibnamefont{and} \bibinfo{author}{\bibfnamefont{F.}~\bibnamefont{{Walter}}},
  \bibinfo{journal}{Mon.\ Not.\ Roy.\ Astron.\ Soc.}
  \textbf{\bibinfo{volume}{340}}, \bibinfo{pages}{12} (\bibinfo{year}{2003}).

\bibitem[{\citenamefont{{Gentile} et~al.}(2004)\citenamefont{{Gentile},
  {Salucci}, {Klein}, {Vergani}, and {Kalberla}}}]{2004MNRAS.351..903G}
\bibinfo{author}{\bibfnamefont{G.}~\bibnamefont{{Gentile}}},
  \bibinfo{author}{\bibfnamefont{P.}~\bibnamefont{{Salucci}}},
  \bibinfo{author}{\bibfnamefont{U.}~\bibnamefont{{Klein}}},
  \bibinfo{author}{\bibfnamefont{D.}~\bibnamefont{{Vergani}}},
  \bibnamefont{and}
  \bibinfo{author}{\bibfnamefont{P.}~\bibnamefont{{Kalberla}}},
  \bibinfo{journal}{Mon.\ Not.\ Roy.\ Astron.\ Soc.}
  \textbf{\bibinfo{volume}{351}}, \bibinfo{pages}{903} (\bibinfo{year}{2004}).

\bibitem[{\citenamefont{{Donato}
  et~al.}(2004{\natexlab{a}})\citenamefont{{Donato}, {Gentile}, and
  {Salucci}}}]{2004MNRAS.353L..17D}
\bibinfo{author}{\bibfnamefont{F.}~\bibnamefont{{Donato}}},
  \bibinfo{author}{\bibfnamefont{G.}~\bibnamefont{{Gentile}}},
  \bibnamefont{and}
  \bibinfo{author}{\bibfnamefont{P.}~\bibnamefont{{Salucci}}},
  \bibinfo{journal}{Mon.\ Not.\ Roy.\ Astron.\ Soc.}
  \textbf{\bibinfo{volume}{353}}, \bibinfo{pages}{L17}
  (\bibinfo{year}{2004}{\natexlab{a}}).

\bibitem[{\citenamefont{{Bahcall} and {Soneira}}(1980)}]{bahcall}
\bibinfo{author}{\bibfnamefont{J.~N.} \bibnamefont{{Bahcall}}}
  \bibnamefont{and} \bibinfo{author}{\bibfnamefont{R.~M.}
  \bibnamefont{{Soneira}}}, \bibinfo{journal}{Astrophys.\ J.\ Suppl.}
  \textbf{\bibinfo{volume}{44}}, \bibinfo{pages}{73} (\bibinfo{year}{1980}).

\bibitem[{\citenamefont{{Donato}
  et~al.}(2004{\natexlab{b}})\citenamefont{{Donato}, {Fornengo}, {Maurin},
  {Salati}, and {Taillet}}}]{pbar_susy}
\bibinfo{author}{\bibfnamefont{F.}~\bibnamefont{{Donato}}},
  \bibinfo{author}{\bibfnamefont{N.}~\bibnamefont{{Fornengo}}},
  \bibinfo{author}{\bibfnamefont{D.}~\bibnamefont{{Maurin}}},
  \bibinfo{author}{\bibfnamefont{P.}~\bibnamefont{{Salati}}}, \bibnamefont{and}
  \bibinfo{author}{\bibfnamefont{R.}~\bibnamefont{{Taillet}}},
  \bibinfo{journal}{Phys.\ Rev.\ D} \textbf{\bibinfo{volume}{69}},
  \bibinfo{pages}{063501} (\bibinfo{year}{2004}{\natexlab{b}}).

\bibitem[{\citenamefont{{Maurin} and {Taillet}}(2003)}]{MaurinTaillet}
\bibinfo{author}{\bibfnamefont{D.}~\bibnamefont{{Maurin}}} \bibnamefont{and}
  \bibinfo{author}{\bibfnamefont{R.}~\bibnamefont{{Taillet}}},
  \bibinfo{journal}{Astron.\ Astroph.} \textbf{\bibinfo{volume}{404}},
  \bibinfo{pages}{949} (\bibinfo{year}{2003}).

\bibitem[{\citenamefont{{Maurin} et~al.}(2003)\citenamefont{{Maurin},
  {Taillet}, {Donato}, {Salati}, {Barrau}, and {Boudoul}}}]{Revue}
\bibinfo{author}{\bibfnamefont{D.}~\bibnamefont{{Maurin}}},
  \bibinfo{author}{\bibfnamefont{R.}~\bibnamefont{{Taillet}}},
  \bibinfo{author}{\bibfnamefont{F.}~\bibnamefont{{Donato}}},
  \bibinfo{author}{\bibfnamefont{P.}~\bibnamefont{{Salati}}},
  \bibinfo{author}{\bibfnamefont{A.}~\bibnamefont{{Barrau}}}, \bibnamefont{and}
  \bibinfo{author}{\bibfnamefont{G.}~\bibnamefont{{Boudoul}}},
  \bibinfo{journal}{Research Signpost}
  \textbf{\bibinfo{volume}{astro-ph/0212111}} (\bibinfo{year}{2003}).

\bibitem[{\citenamefont{{Maurin} et~al.}(2001)\citenamefont{{Maurin}, {Donato},
  {Taillet}, and {Salati}}}]{Usine1}
\bibinfo{author}{\bibfnamefont{D.}~\bibnamefont{{Maurin}}},
  \bibinfo{author}{\bibfnamefont{F.}~\bibnamefont{{Donato}}},
  \bibinfo{author}{\bibfnamefont{R.}~\bibnamefont{{Taillet}}},
  \bibnamefont{and} \bibinfo{author}{\bibfnamefont{P.}~\bibnamefont{{Salati}}},
  \bibinfo{journal}{Astrophys.\ J.} \textbf{\bibinfo{volume}{555}},
  \bibinfo{pages}{585} (\bibinfo{year}{2001}).

\bibitem[{\citenamefont{{Taillet} and {Maurin}}(2003)}]{TailletMaurin}
\bibinfo{author}{\bibfnamefont{R.}~\bibnamefont{{Taillet}}} \bibnamefont{and}
  \bibinfo{author}{\bibfnamefont{D.}~\bibnamefont{{Maurin}}},
  \bibinfo{journal}{Astron.\ Astroph.} \textbf{\bibinfo{volume}{402}},
  \bibinfo{pages}{971} (\bibinfo{year}{2003}).

\bibitem[{\citenamefont{{Jones}}(1978)}]{1978ApJ...222.1097J}
\bibinfo{author}{\bibfnamefont{F.~C.} \bibnamefont{{Jones}}},
  \bibinfo{journal}{Astrophys.\ J.} \textbf{\bibinfo{volume}{222}},
  \bibinfo{pages}{1097} (\bibinfo{year}{1978}).

\bibitem[{\citenamefont{{Donato} et~al.}(2002)\citenamefont{{Donato}, {Maurin},
  and {Taillet}}}]{2002A&A...381..539D}
\bibinfo{author}{\bibfnamefont{F.}~\bibnamefont{{Donato}}},
  \bibinfo{author}{\bibfnamefont{D.}~\bibnamefont{{Maurin}}}, \bibnamefont{and}
  \bibinfo{author}{\bibfnamefont{R.}~\bibnamefont{{Taillet}}},
  \bibinfo{journal}{Astron.\ Astroph.} \textbf{\bibinfo{volume}{381}},
  \bibinfo{pages}{539} (\bibinfo{year}{2002}).

\bibitem[{\citenamefont{{Combet} et~al.}(2005)\citenamefont{{Combet}, {Maurin},
  {Donnelly}, {O'C.~Drury}, and {Vangioni-Flam}}}]{celine}
\bibinfo{author}{\bibfnamefont{C.}~\bibnamefont{{Combet}}},
  \bibinfo{author}{\bibfnamefont{D.}~\bibnamefont{{Maurin}}},
  \bibinfo{author}{\bibfnamefont{J.}~\bibnamefont{{Donnelly}}},
  \bibinfo{author}{\bibfnamefont{L.}~\bibnamefont{{O'C.~Drury}}},
  \bibnamefont{and}
  \bibinfo{author}{\bibfnamefont{E.}~\bibnamefont{{Vangioni-Flam}}},
  \bibinfo{journal}{Astron.\ Astroph.} \textbf{\bibinfo{volume}{435}},
  \bibinfo{pages}{151} (\bibinfo{year}{2005}).

\bibitem[{\citenamefont{{Bringmann}}(2005)}]{torsten}
\bibinfo{author}{\bibfnamefont{T.}~\bibnamefont{{Bringmann}}},
  \textbf{\bibinfo{volume}{arXiv:astro-ph/0506219}} (\bibinfo{year}{2005}).

\bibitem[{\citenamefont{{Pochon}}(2005)}]{jonathan}
\bibinfo{author}{\bibfnamefont{J.}~\bibnamefont{{Pochon}}},
  \bibinfo{journal}{Ph.D. thesis - Chapter 9}  (\bibinfo{year}{2005}).

\bibitem[{\citenamefont{\mbox{HESS Collaboration -- F. Aharonian {\it et
  al.}}}(2004)}]{hess}
\bibinfo{author}{\bibnamefont{\mbox{HESS Collaboration -- F. Aharonian {\it et
  al.}}}}, \bibinfo{journal}{Astron.\ Astroph.} \textbf{\bibinfo{volume}{425}},
  \bibinfo{pages}{L13} (\bibinfo{year}{2004}).

\bibitem[{\citenamefont{{Diemand} et~al.}(2005)\citenamefont{{Diemand},
  {Moore}, and {Stadel}}}]{mini_clump}
\bibinfo{author}{\bibfnamefont{J.}~\bibnamefont{{Diemand}}},
  \bibinfo{author}{\bibfnamefont{B.}~\bibnamefont{{Moore}}}, \bibnamefont{and}
  \bibinfo{author}{\bibfnamefont{J.}~\bibnamefont{{Stadel}}},
  \bibinfo{journal}{Nature} \textbf{\bibinfo{volume}{433}},
  \bibinfo{pages}{389} (\bibinfo{year}{2005}).

\bibitem[{\citenamefont{\mbox{HEAT Collaboration -- S.~W. Barwick {\it et
  al.}}}(1997)}]{heat_1}
\bibinfo{author}{\bibnamefont{\mbox{HEAT Collaboration -- S.~W. Barwick {\it et
  al.}}}}, \bibinfo{journal}{Astrophys.\ J.} \textbf{\bibinfo{volume}{482}},
  \bibinfo{pages}{L191} (\bibinfo{year}{1997}).

\bibitem[{\citenamefont{\mbox{S. Coutu {\it et al.}}}(1999)}]{heat_2}
\bibinfo{author}{\bibnamefont{\mbox{S. Coutu {\it et al.}}}},
  \bibinfo{journal}{Astropart.\ Phys.} \textbf{\bibinfo{volume}{11}},
  \bibinfo{pages}{429} (\bibinfo{year}{1999}).

\end{thebibliography}

%
\end{document}